\documentclass[longauth,bibyear]{aa}
\begin{document}
\title{The sharp ALMA view of infall and outflow in the massive protocluster G31.41+0.31}
\author{M.\ T.\ Beltr\'an\inst{1}, V.\ M.\ Rivilla\inst{2, 1}, R.\ Cesaroni\inst{1}, D.\ Galli\inst{1}, L.\ Moscadelli\inst{1}, 
A.\ Ahmadi\inst{3}, 
H.\ Beuther\inst{4}, 
S.\ Etoka\inst{5}, 
C.\ Goddi\inst{6, 7}, 
P.\ D.\ Klaassen\inst{8}, 
R.\ Kuiper\inst{9}, 
M.\ S.\ N.\ Kumar\inst{10}, 
A.\ Lorenzani\inst{1}, 
T.\ Peters\inst{11}, 
\'A.\ S\'anchez-Monge\inst{12}, 
P.\ Schilke\inst{12}, 
F. van der Tak\inst{13, 14}, 
S.\ Vig\inst{15}
}
\institute{
INAF-Osservatorio Astrofisico di Arcetri, Largo E.\ Fermi 5,
I-50125 Firenze, Italy
\and
Centro de Astrobiolog\'{i}a (CSIC-INTA), Ctra.\ de Ajalvir Km. 4, Torrej\'on de Ardoz, 28850 Madrid, Spain
\and
Leiden Observatory, Leiden University, PO Box 9513, 2300 RA Leiden, The Netherlands 
\and
 Max-Planck-Institut f\"ur Astronomie, K\"onigstuhl 17, 69117 Heidelberg, Germany
\and
Jodrell Bank Centre for Astrophysics, The University of Manchester, Alan Turing Building, Manchester M13 9PL, UK
\and
Dipartimento di Fisica, Universit\`a degli Studi di Cagliari, SP Monserrato-Sestu km 0.7, I-09042 Monserrato, Italy
\and
INAF-Osservatorio Astronomico di Cagliari, via della Scienza 5, I-09047 Selargius, Italy
\and
UK Astronomy Technology Centre, Royal Observatory Edinburgh, Blackford Hill, Edinburgh EH9 3HJ, UK
\and
Zentrum f\"ur Astronomie der Universit\"at Heidelberg, Institut f\"ur Theoretische Astrophysik, Albert-Ueberle-Stra{\ss}e 2, 69120 Heidelberg, Germany
\and
Instituto de Astrof\'{i}sica e Ci\^encias do Espa\c{c}o, Universidade do Porto, CAUP, Rua das Estrelas, 4150-762, Porto, Portugal
\and
Max-Planck-Institut f\"{u}r Astrophysik, Karl-Schwarzschild-Str. 1, D-85748 Garching, Germany
\and
I.\ Physikalisches Institut, Universit\"at zu K\"oln, Z\"ulpicher Str. 77, 50937 K\"oln, Germany
\and
Kapteyn Astronomical Institute, University of Groningen, 9700 AV Groningen, The Netherlands
\and
SRON Netherlands Institute for Space Research, Landleven 12, 9747 AD Groningen, The Netherlands
\and
Department of Earth and Space Sciences, Indian Institute of Space Science and Technology, Thiruvananthapuram, 695 547, India
}
\offprints{M.\ T.\ Beltr\'an, \email{maria.beltran@inaf.it}}

\date{Received date; accepted date}

\titlerunning{Infall and outflow in G31.41+0.31}
\authorrunning{Beltr\'an et al.}

\abstract
{To better understand the formation of high-mass stars, it is fundamental to investigate how matter accretes onto young massive stars, how it is ejected, and how all this differs from the low-mass case. The massive protocluster G31.41+0.31 is the ideal target to study all these processes because observations at millimeter and centimeter wavelengths have resolved the emission of the {\it Main} core into at least four massive dust continuum sources, named A, B, C, and D,  within 1$''$ or 0.018\,pc, and have identified signatures of infall and several outflows associated with the core.}
{We study the interplay between infall and outflow in G31.41+0.31 by investigating at a spatial resolution of a few 100\,au their properties  and their possible impact on the core.}
{We carried out molecular line observations of typical high-density tracers, such as CH$_3$CN or H$_2$CO, and shock and outflow tracers, such as SiO, with ALMA at 1.4\,mm that achieved an angular resolution of $\sim$0$\farcs$09 ($\sim$340\,au).}
{The observations have revealed inverse P-Cygni profiles in CH$_3$CN and H$_2$CO toward the four sources embedded in the {\it Main} core, suggesting that all of them are undergoing collapse. The infall rates, estimated from the red-shifted absorption are on the order of  $\sim$10$^{-2}$\,$M_\odot$\,yr$^{-1}$. The individual infall rates imply that the accretion timescale of the {\it Main} core is an order of magnitude smaller than its rotation timescale. This confirms that rotating toroids such as the G31 {\it Main} core are non-equilibrium, transient collapsing structures that need to be constantly replenished with fresh material from a large-scale reservoir. For sources B, C, and D, the infall could be accelerating inside the sources, while for source A, the presence of a second emission component complicates the interpretation. The SiO observations have revealed the presence of at least six outflows in the G31.41+0.31 star-forming region, and suggest that each of the four sources embedded in the {\it Main} core drives a molecular outflow. The outflow rates are on the order of $\sim$10$^{-5}$--10$^{-4}\,M_\odot$\,yr$^{-1}$, depending on the SiO abundance. The feedback of the outflows appears sufficient to sustain the turbulence in the core and to eventually disrupt the core or prevent its further collapse. The mass accretion rates onto the individual sources, estimated from the highest values of the outflow mass loss rates, are on the order of 10$^{-4}$\,$M_\odot$\,yr$^{-1}$. The difference of about 2 orders of magnitude between the accretion rates and the infall rates suggests that the central stars and the jets/outflows might not be efficient in removing disk material, which could lead to gravitational instabilities, fragmentation of the disk, and episodic accretion events.} 
{Infall and outflows are simultaneously present in all four sources embedded in the high-mass star-forming core G31.41+0.31. This indicates that these sources are still actively accreting and have not reached their final mass yet, which confirms the youth of this massive core.}
\keywords{ISM: individual objects: G31.41+0.31 
-- stars: formation -- stars: massive -- techniques: interferometric}

\maketitle

\section{Introduction}
\label{sect-intro}

The formation of stars with masses $>$20--30\,$M_\sun$ is among the least understood phenomena in modern astrophysics. Observations of massive outflows (e.g., Shepherd \& Churchwell~\cite{shepherd96}; Arce et al.~\cite{arce07}) suggest a scaled up version of low-mass star formation (Krumholz~\cite{krumholz15}), but the lack of evidence for massive pre-stellar cores (Motte et al.~\cite{motte18}) and, in particular, of protostellar disks around very massive (early O-type) stars (Beltr\'an \& de Wit~\cite{beltran16}; Beltr\'an~\cite{beltran20}) confronts that idea. 

One way to investigate the possible similarities between low-mass and high-mass star formation is to study outflow and infall associated with young stellar objects. According to theory, the accretion and ejection of material are two phenomena closely related in the formation of stars (e.g., Shu et al.~\cite{shu87}), and this relationship has been observationally established for low-mass stars (e.g., Frank et al.~\cite{frank14},  and references therein). However, while the presence of jets and outflows is easily identifiable thanks to their large-scale emission and to their characteristic line profiles with high-velocity wings, the evidence for collapse is more difficult to reveal. This is in part due to the fact that infall and accretion usually happen in the inner regions of the cores, where infall motions can be confused with rotation and outflow. The situation is particularly complicated for high-mass young stellar objects, because the high visual extinction in massive star-forming cores makes their central regions usually only detectable at far-infrared and longer wavelengths. Red-shifted molecular absorption against a bright background continuum source is the only unambiguous way of diagnosing infall, because the absorbing material must be located in front of the protostar (e.g., Chandler
et al.~\cite{chandler05}; Wyroswki et al.~\cite{wyrowski12}). Infall motions are easily distinguishable from outflow and rotation, because in the former case, one should see blue-shifted absorption or possibly emission, while in the latter case, since the motions are perpendicular to the line of sight, any absorption should occur only at systemic velocities.

To study infall in high-mass star-forming regions, observations of high spatial resolution and sensitivity of high-density tracers with excitation energies below the brightness temperature of the central continuum source are required. In this way it is possible to trace the gas of the inner parts of these distant and complex star-forming regions. The advent of (sub)millimeter interferometers has revolutionized the field, as it has allowed for the first time to study the dust close to the embedded massive protostar(s) and the kinematics of the gas at (sub)arcsecond resolution (e.g., Olmi et al.~\cite{olmi96}; Wyrowski et al.~\cite{wyrowski99}, \cite{wyrowski02}), highlighting the occurrence of infall (e.g., Ho \& Haschick~\cite{ho86}; Keto et al.~\cite{keto88}). In fact, in recent years, evidence of collimated outflows and infall signatures have been identified in massive young stellar objects (e.g., Beuther et al.~\cite{beuther02}; Arce et al.~\cite{arce07}; Beltr\'an~\cite{beltran11}, and references therein), establishing also for high-mass stars the relationship between accretion and ejection. However, the number of studies of infall and outflows in high-mass star-forming regions at scales of a few 100s of au is still limited (e.g., Goddi et al.~\cite{goddi20}), in particular if compared to those of the low-mass regime.

 To characterize the motions of the deeply embedded gas that could reveal signatures of rotation and infall and would help us to understand the formation of massive stars, for many years we have carried out multiple high-angular resolution studies of hot molecular cores (HMCs), first with the IRAM Plateau de Bure (PdB) and Submillimeter Array (SMA) interferometers (Beltr\'an et al.~\cite{beltran04}, \cite{beltran05}, \cite{beltran11a}, \cite{beltran11b}), and later on with the Atacama Large Millimeter/submillimeter Array (ALMA) (Cesaroni et al.~\cite{cesa17}; Beltr\'an et al.~\cite{beltran18}, \cite{beltran21}; Moscadelli et al.~\cite{mosca18}, \cite{mosca21}; Goddi et al.~\cite{goddi20}) and the IRAM NOrthern Extended Millimeter Array (NOEMA) (Gieser et al.~\cite{gieser19}, \cite{gieser21}). Cesaroni et al.~(\cite{cesa17}) carried out an extensive and systematic search for circumstellar disks around O-type stars by observing a sample of HMCs at 1.4\,mm and $0\farcs2$ resolution with ALMA. One of the HMCs of the sample is G31.41+0.31 (G31 hereafter), which is located at 3.75 kpc (Immer et al.~\cite{immer19}), has a luminosity of $\sim$5$\times10^4\,L_\sun$ (Osorio et al.~\cite{osorio09}), and displays a clear NE--SW velocity gradient suggestive of rotation (e.g., Beltr\'an et al.~\cite{beltran04}). Among the HMCs studied, G31 stands out as the only source that displays accelerating infall and rotational spin-up (Beltr\'an et al.~\cite{beltran18}). This characteristic suggests the source to be in an earlier evolutionary stage compared to all other targets in the Cesaroni et al.\ sample. At an angular resolution of $0\farcs2$ ($\sim$750\,au), the dust continuum emission traces a well resolved, monolithic, and featureless core, called {\it Main} by Beltr\'an et al.~(\cite{beltran18}), with no hint of fragmentation, despite the high mass ($\sim$70\,$M_\sun$; Cesaroni et al.~\cite{cesa19}) and the diameter of $\sim$8000\,au. However, new ALMA observations at 1.4 and 3.5\,mm at a higher angular resolution of $\sim$0$\farcs1$ ($\sim$375\,au), and Very Large Array (VLA) observations at 7\,mm and $\sim$0$\farcs$05  ($\sim$190\,au) resolution have resolved for the first time the HMC into a small protocluster composed of at least four massive sources, named A, B, C, and D and with masses ranging from $\sim$15 to $\sim$26\,$M_\sun$, within the central $1''$ ($\sim$3750\,au) region of the core (Beltr\'an et al.~\cite{beltran21}). These observations have revealed that the homogeneous appearance previously observed at 1.4\,mm and $0\farcs2$ is a consequence of both high dust opacity and insufficient angular resolution. Besides the four sources embedded in the {\it Main} core, there are six additional millimeter sources located very close to the core which appear to outline streams/filaments of matter pointing to the HMC (Beltr\'an et al.~\cite{beltran21}).

The four dust continuum sources embedded in the {\it Main} core have also been detected at centimeter wavelengths (Cesaroni et al.~\cite{cesa10}; Beltr\'an et al.~\cite{beltran21}). The analysis of their spectral energy distribution suggests that the centimeter emission of sources A, B, and D probably originates from thermal radio jets. This would indicate the presence of different molecular outflows associated with the sources embedded in the core and, in fact, CO and SiO observations have revealed the presence of several outflows in G31 (Olmi et al.~\cite{olmi96}; Cesaroni et al.~\cite{cesa11}; Beltr\'an et al.~\cite{beltran18}).

The massive collapsing protocluster G31 is the ideal target to investigate the properties of infall and outflow and their  relationship in the high-mass star-forming regions at scales of a few 100s of au. In this study,  we analyze molecular line observations carried out at 1.4\,mm and $\sim$0$\farcs09$ resolution ($\sim$340\,au) with ALMA to trace the kinematics in the central region of this high-mass star-forming core. The spectral setup of the interferometer includes typical high-density tracers, such as CH$_3$CN and isotopologues and H$_2$CO, and the shock and outflow tracer SiO. The aim of the observations is to investigate the infall toward the individual sources embedded in the {\it Main} core and measure the corresponding mass infall rates from the infall velocity and the gas column density, as well as to study the properties of the molecular outflows detected in the region and their possible impact on the star formation in the G31 core.

This article is organized as follows: in Sect.~2 we describe the
observations; in Sect.~3 we present the results, the emission of dense core and molecular outflow tracers; in Sect.~4 we analyze the properties of the molecular outflows; in Sect.~5 we discuss the infall toward the embedded sources. Finally, in Sect.~5  we give our main conclusions.
 

\begin{figure*}
\centerline{\includegraphics[angle=0,width=17cm,angle=0]{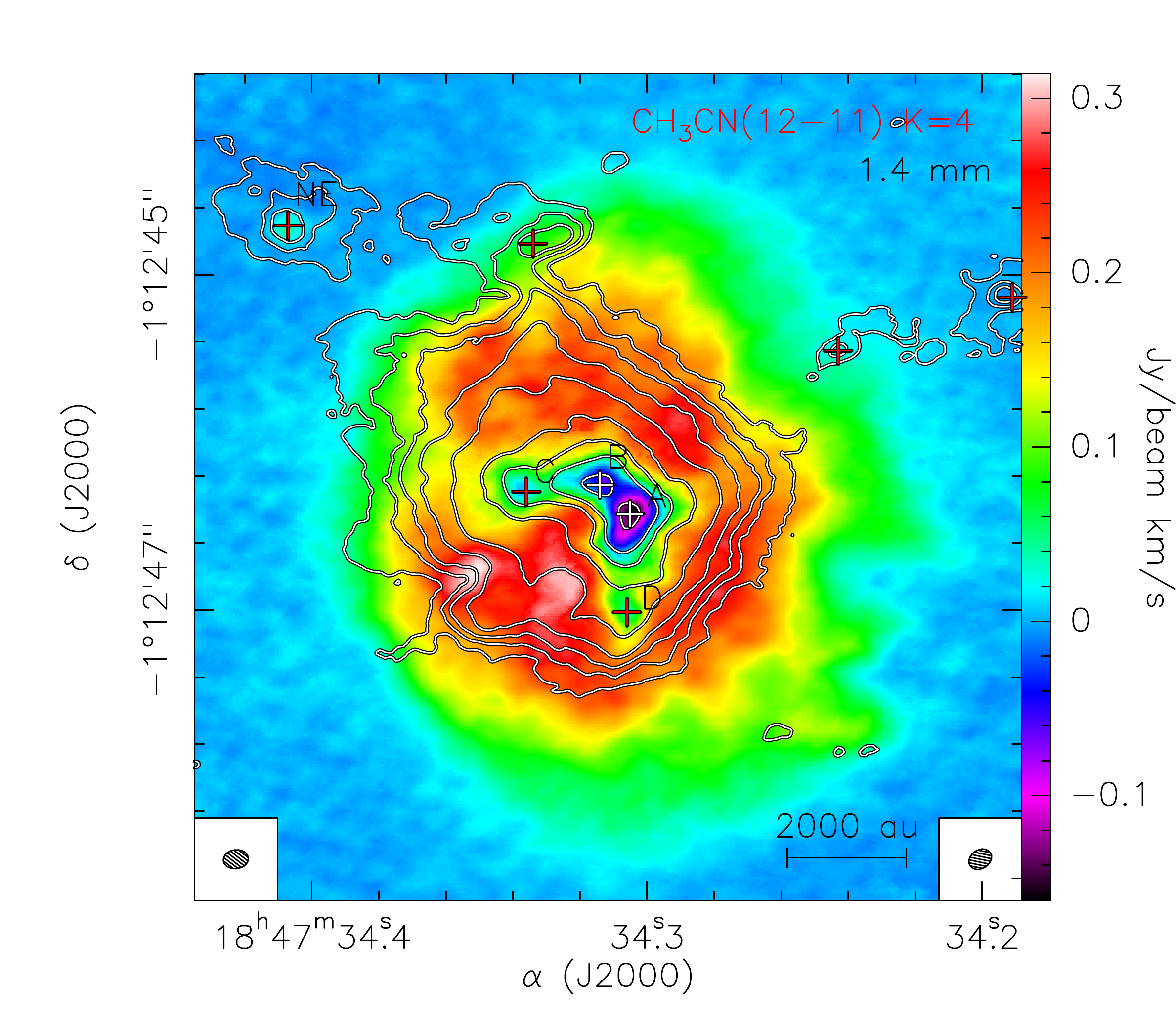}}
\caption{Overlay of the ALMA 1.4\,mm continuum emission ({\it white contours}) of G31.41+0.31 on the integrated intensity (moment 0) map ({\it colors}) of CH$_3$CN\, $K = 4$ ($E_{\rm up}/k_{\rm B}$=183\,K). The line emission has been integrated over the velocity range 92 to 102\,km\,s$^{-1}$. The contours are  3, 6, 9, 15, 30, 60, 90, 120, and 165 times 1\,$\sigma$, which is 0.32\,mJy\,beam$^{-1}$. The red crosses indicate the positions of the sources detected in the region by Beltr\'an et al.~(\cite{beltran21}). The synthesized beam of the 1.4\,mm continuum emission ($0\farcs11\times0\farcs08$, P.A.\ =$-83^\circ$) and CH$_3$CN\, $K = 4$ ($0\farcs11\times0\farcs08$, P.A.\ =$-57^\circ$) are shown in the lower left- and right-hand corner, respectively.}
\label{fig-k4}
\end{figure*}

\begin{figure*}
\centerline{\includegraphics[angle=0,width=17cm,angle=0]{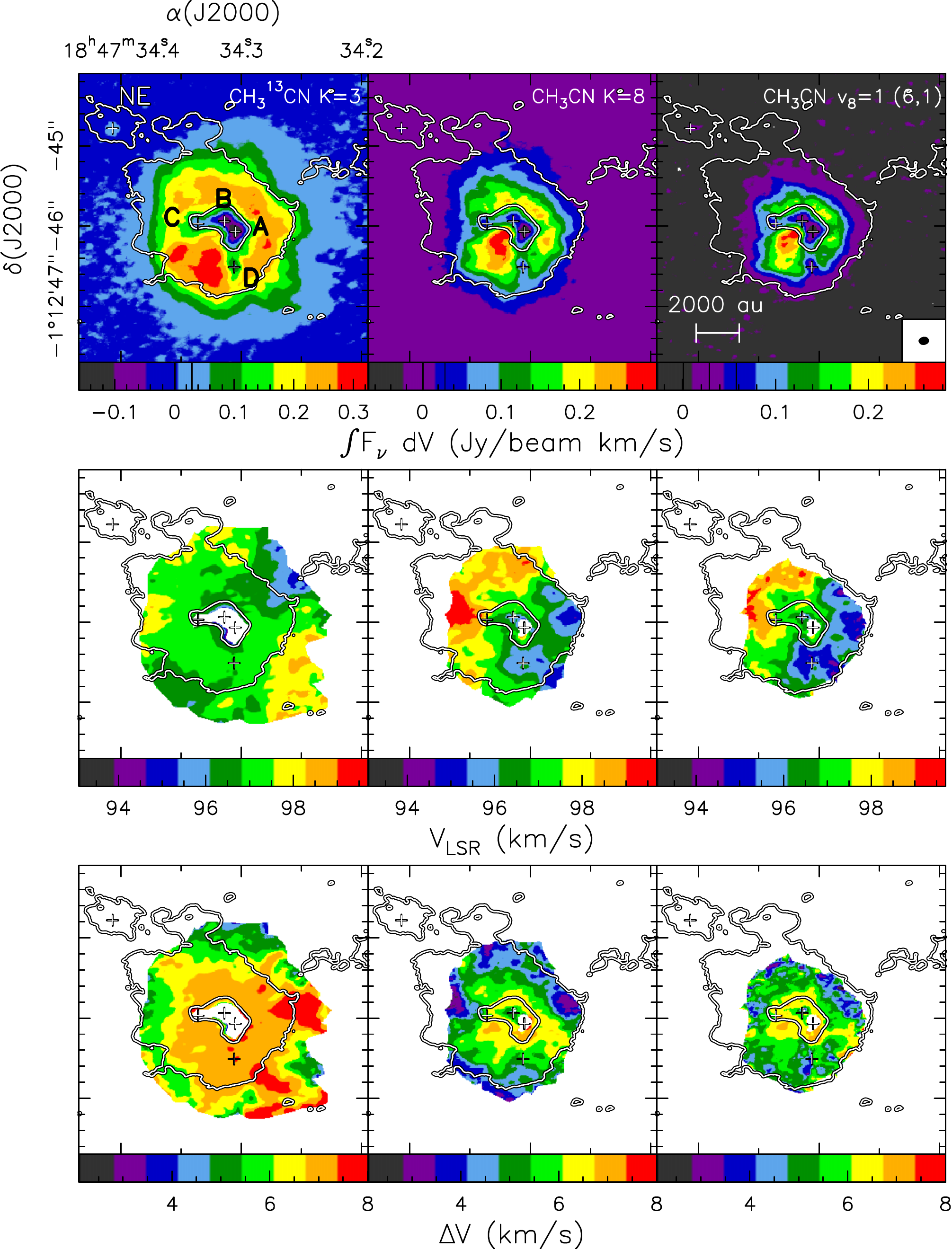}}
\caption{({\it Top panels}) Overlay of the ALMA 1.4\,mm continuum emission ({\it contours}) on the integrated intensity (moment 0) map ({\it colors}) of CH${_3}^{13}$CN\ $K=3$ ($E_{\rm up}/k_{\rm B}$=142\,K), CH$_3$CN\ $K=8$ ($E_{\rm up}/k_{\rm B}$=526\,K), and CH$_3$CN\  $K, l=(6, 1)$ $v_8=1$ ($E_{\rm up}/k_{\rm B}$=778\,K) (12--11). The contours correspond to the 3$\sigma$ and 50\% levels. ({\it Middle panels}) Line velocity (moment 1) maps for the same molecular species. ({\it  Bottom panels}) Velocity dispersion (moment 2) maps for the same species. The moment maps have been integrated over the velocity range 92 to 102\,km\,s$^{-1}$. The white crosses indicate the positions of the $NE$ core and of the four continuum sources embedded in the $Main$ core detected by Beltr\'an et al.~(\cite{beltran21}). The synthesized beam is shown in the lower right-hand corner of the top right panel.}
\label{fig-moments}
\end{figure*}


\section{Observations}
\label{sect-ALMA}

Interferometric observations of G31 at 1.4\,mm in Band 6 were carried out with ALMA in Cycle 4 as part of project 2016.1.00223.S (P.I.: M.\ Beltr\'an). The observations were carried out in one of the most extended configurations, C40-7, and were combined with those of the ALMA Cycle 2 project 2013.1.00489.S (P.I.: R. Cesaroni), which were observed in the extended C34-7/(6) configuration and with the same digital correlator configuration: thirteen spectral windows (SPW) that cover, among other lines, SiO\,(5--4), H$_2$CO\,(3$_{0,3}$--2$_{0,2}$), (3$_{2,2}$--2$_{2,1}$), and (3$_{2,1}$--2$_{2,0}$), CH$_3$CN\,(12$_K$--11$_K$), CH$_3^{13}$CN\,(12$_K$--11$_K$), CH$_3$CN\,$v_8$ = 1\,(12$_K$--11$_K$), and $^{13}$CH$_3$CN\,(13$_K$--12$_K$). The SPW0 window that covers SiO and three transitions of H$_2$CO has a bandwidth of 1875\,MHz and a spectral resolution of 1.95 MHz, which corresponds to a velocity resolution of $\sim$2.7\,km\,s$^{-1}$, and was used to determine the continuum emission. The spectral resolution of CH$_3$CN and CH$_3^{13}$CN is 0.244\,MHz, which translates into a velocity resolution of $\sim$0.33\,km\,s$^{-1}$, and that of CH$_3$CN\,$v_8$ and $^{13}$CH$_3$CN is 0.488\,MHz, which corresponds to $\sim$0.66\,km\,s$^{-1}$. We refer to Cesaroni et al.~(\cite{cesa17}), Beltr\'an et al.~(\cite{beltran18}), and  Beltr\'an et al.~(\cite{beltran21}) for detailed information on the observations. 

The phase reference center of the observations was set to the position $\alpha$(J2000) = 18$^{\rm h}$ 47$^{\rm m}$ 34$\fs$315, $\delta$(J2000) = $-$01$^\circ$ 12$'$ 45$\farcs$9.  The data were calibrated and imaged using the {\sc CASA}\footnote{The {\sc CASA} package is available at \url{http://casa.nrao.edu/}} software package. We assumed the standard 
uncertainty on the ALMA absolute flux calibration of $\sim$10\% in Band 6. Maps were created with the ROBUST parameter of Briggs~(\cite{briggs95}) set equal to 0. Further imaging and analysis were done with the {\sc GILDAS}\footnote{The {\sc GILDAS} package is available at \url{http://www.iram.fr/IRAMFR/GILDAS}} software package. The continuum was subtracted from the line emission using the {\sc STATCONT}\footnote{\url{http://www.astro.uni-koeln.de/~sanchez/statcont}} algorithm (S\'anchez-Monge et al.~\cite{sanchez-monge18}). The resulting synthesized CLEANed beam of the maps is $0\farcs11\times0\farcs08$,  which corresponds to spatial scales of $\sim$350\,au at the distance of G31, and position angle, P.A.,  $-83^\circ$ for SiO (5--4) and H$_2$CO (3$_{0,3}$--2$_{0,2}$), (3$_{2,2}$--2$_{2,1}$), and (3$_{2,1}$--2$_{2,0}$), $-57^\circ$ for CH$_3$CN (12--11) $K=0$ to 7 and CH$_3^{13}$CN (12--11) $K=0$ to 4, $+73^\circ$ for CH$_3$CN (12--11) $K\ge8$ and CH$_3^{13}$CN (12--11) $K\geq5$,  $-79^\circ$ for $^{13}$CH$_3$CN (13--12),  and $-32^\circ$ for CH$_3$CN $v_8=1$ (12--11). The 
rms noise of the maps is 0.45\,mJy\,beam$^{-1}$ for SiO and H$_2$CO, and 0.8--0.9\,mJy\,beam$^{-1}$ for CH$_3$CN, CH$_3^{13}$CN,  CH$_3$CN\,$v_8$ = 1, and $^{13}$CH$_3$CN.

\begin{figure*}
\centerline{\includegraphics[angle=0,width=17cm,angle=0]{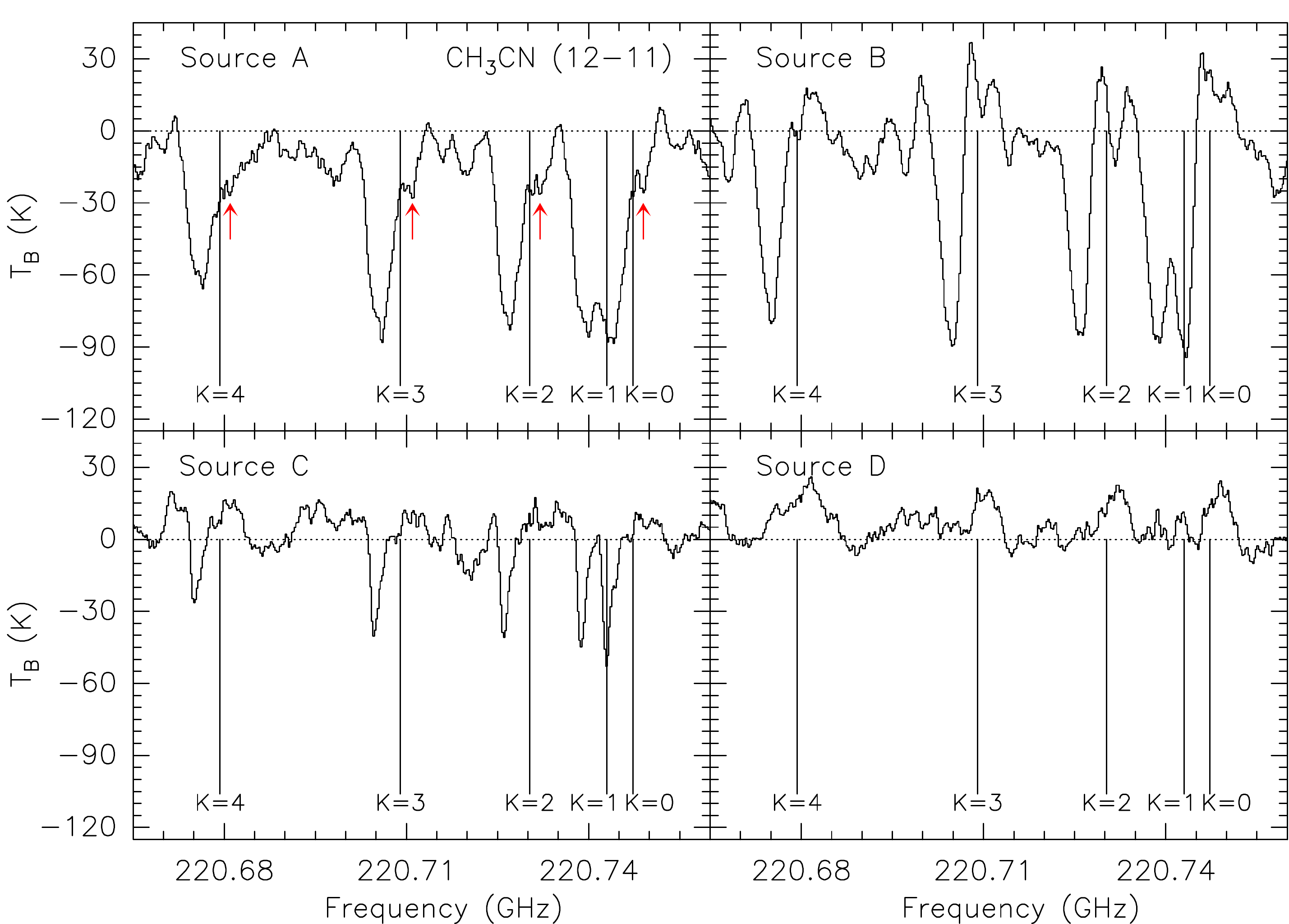}}
\caption{CH$_3$CN\,(12--11) spectra of the $K=0, 1, 2, 3$ and 4 transitions averaged over a beam toward the position of the dust continuum emission peak of the four sources embedded in the {\it Main} core of G31. The black lines  indicate the different K components shifted to the frequency of the ambient gas velocity (96.5\,km\,s$^{-1}$). The red arrows in the top left panel indicate the secondary absorption component at blue-shifted velocities associated with source A.}
\label{fig-spw3}
\end{figure*} 

\section{Results and Analysis}

The configuration of the digital correlator of ALMA allows to observe a wide range of frequencies simultaneously and, for sources as chemically rich as G31, to observe the emission of species tracing different physical conditions (temperature and density) and phenomena. In this work, we have focused on CH$_3$CN\ (ground state and vibrationally excited), SiO, and H$_2$CO to trace infall and outflow in the region. 

\subsection{Infall emission}
\label{sect-ch3cn}

Beltr\'an et al.~(\cite{beltran18}) analyzed 1.4\,mm ALMA observations toward the {\it Main} core of the HMC G31 carried out at an angular resolution of $\sim$0$\farcs$2 ($\sim$750\,au). These observations have revealed that the high-density tracers, such as CH$_3$CN and CH$_3$OCHO, trace a ring-like structure while the dust continuum traces a homogeneous and almost spherical core whose emission peaks at the central dip of the line emission. This morphology has been observed for transitions with very different excitation conditions and upper level energies, which range from $\sim$70\,K to 850\,K. The spectra of these transitions toward the dust continuum emission peak have clearly shown inverse P-Cygni profiles, with strong red-shifted absorption even for transitions with high upper level excitation energies, $E_{\rm up}/k_{\rm B}$,  such as CH$_3$CN $K=8$ ($E_{\rm up}/k_{\rm B}$=526\,K) and CH$_3$CN $v_8$=1 $K,l$=(6, 1) ($E_{\rm up}/k_{\rm B}$=778\,K).

Here, the new ALMA observations at a higher angular resolution of $\sim$0$\farcs09$ ($\sim$340\,au) confirm the presence of absorption toward the center of the {\it Main} core of G31 and show that this absorption is indeed detected toward all the four sources, A, B, C, and D, embedded in it. This is clearly seen for sources A and B in Fig.~\ref{fig-k4}, where the 1.4\,mm dust continuum emission is overlaid on the integrated intensity map of the CH$_3$CN $K=4$ transition, and for all sources in Fig.~\ref{fig-moments}, which shows the moment maps for CH$_3^{13}$CN $K=3$, CH$_3$CN $K=8$, and CH$_3$CN $v_8$=1 $K,l$=(6, 1). As seen in these figures, the integrated line emission clearly decreases toward the position of the four continuum sources, and  for sources A and B, the moment\,0 maps show absorption. In CH$_3^{13}$CN $K=3$ (Fig.~\ref{fig-moments}), the absorption toward the center of the {\it Main} core outlines the 50\% contour level of the  dust continuum emission at 1.4\,mm. 

\subsubsection{Moment maps}

As already mentioned, the moment 0 maps of CH$_3$CN\ and isotopologues  (Fig.~\ref{fig-moments}) clearly indicate the presence of absorption toward the center of the G31 {\it Main} core. Besides this central absorption, the integrated intensity maps also show an increase in emission to the southeast of sources A and B, which becomes spatially smaller and approaches the sources as the upper level energy of the transition ($E_{\rm up}/k_{\rm B}$) increases.  The enhancement of methyl cyanide emission coincides with the direction of an SiO molecular outflow powered by source B (see Sect.~\ref{sect-indi-out}), and this suggests that the gas-phase abundance of CH$_3$CN might have been locally enhanced in protostellar shocks (e.g. Arce et al.~\cite{arce08}; Codella et al.~\cite{codella09}; Palau et al.~\cite{palau17}).

The line velocity maps trace the NE--SW velocity gradient  seen in different species and interpreted as rotation of the core by different studies (e.g., Beltr\'an et al.~\cite{beltran04}, \cite{beltran05}, \cite{beltran18}; Girart et al.~\cite{girart09}; Cesaroni et al.~\cite{cesa11}). This velocity gradient is less evident in CH$_3^{13}$CN $K=3$, but this could be a result of the interaction of the core with the multiple outflows in the area (see Sect.~\ref{sect-indi-out}). The velocity dispersion maps show an increase toward the central position, in particular to south-east of source B and to west and south-west of source A, in correspondence with the SiO molecular outflows associated with these sources as discussed in Sect.~\ref{sect-indi-out}. The enhancement associated with the outflow powered by source A is more evident in the moment\,2 map of CH$_3^{13}$CN $K=3$.

\begin{figure}
\centerline{\includegraphics[angle=0,width=8cm,angle=0]{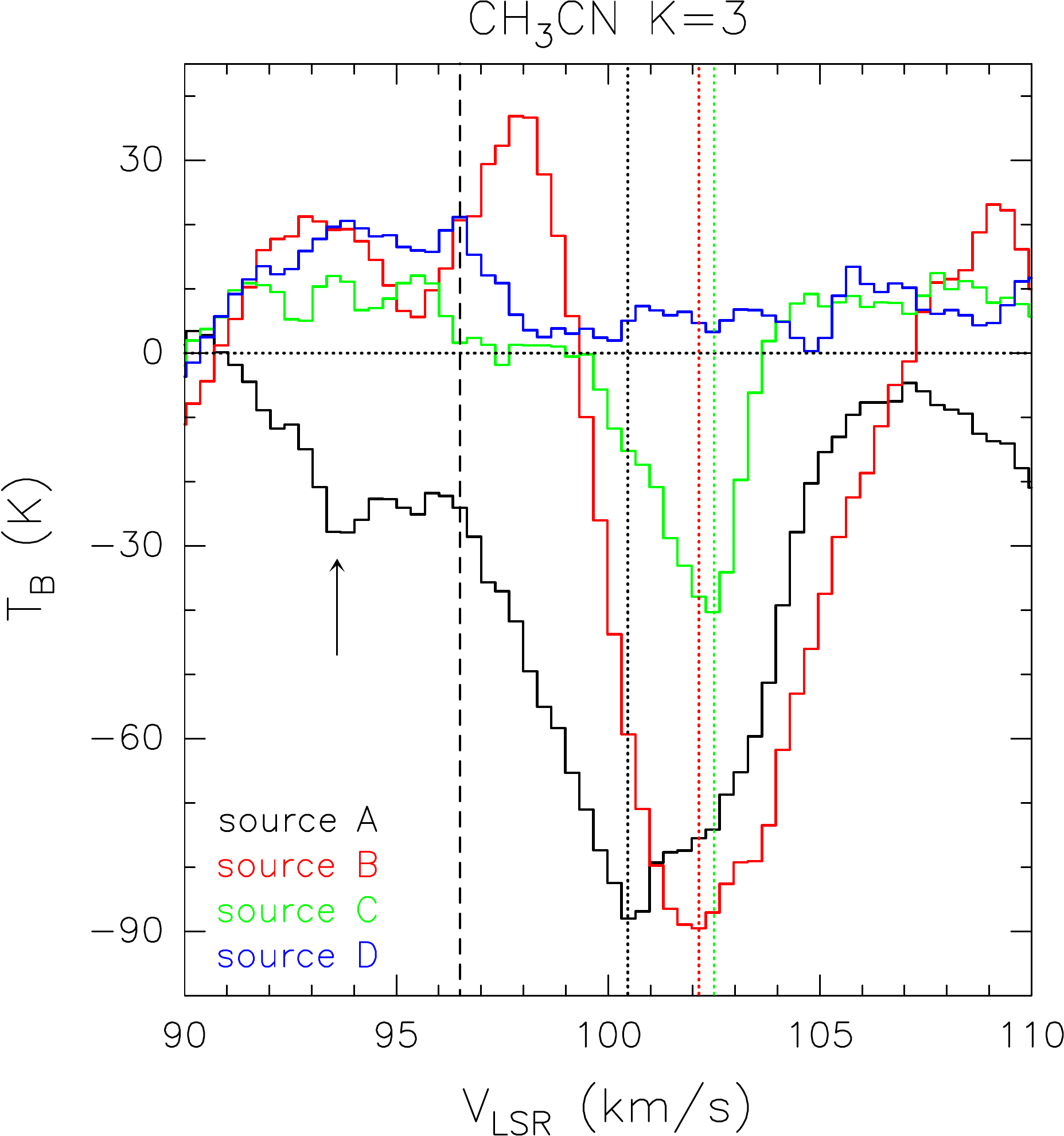}}
\caption{Same as Fig.~\ref{fig-spw3} but for the CH$_3$CN\,(12--11) $K=3$ transition. The $x$-axis is in units of velocity.  The vertical dashed line indicates the systemic LSR velocity of 96.5$\pm0.5$\,km\,s$^{-1}$ (Beltr\'an et al.~\cite{beltran18}). The black, red, and green dotted vertical lines indicate the velocity of the deepest absorption toward sources A, B, and C, respectively. The black arrow indicates the secondary absorption component at blue-shifted velocities associated with source A.}
\label{fig-k3}
\end{figure}

\begin{figure}
\centerline{\includegraphics[angle=0,width=8cm,angle=0]{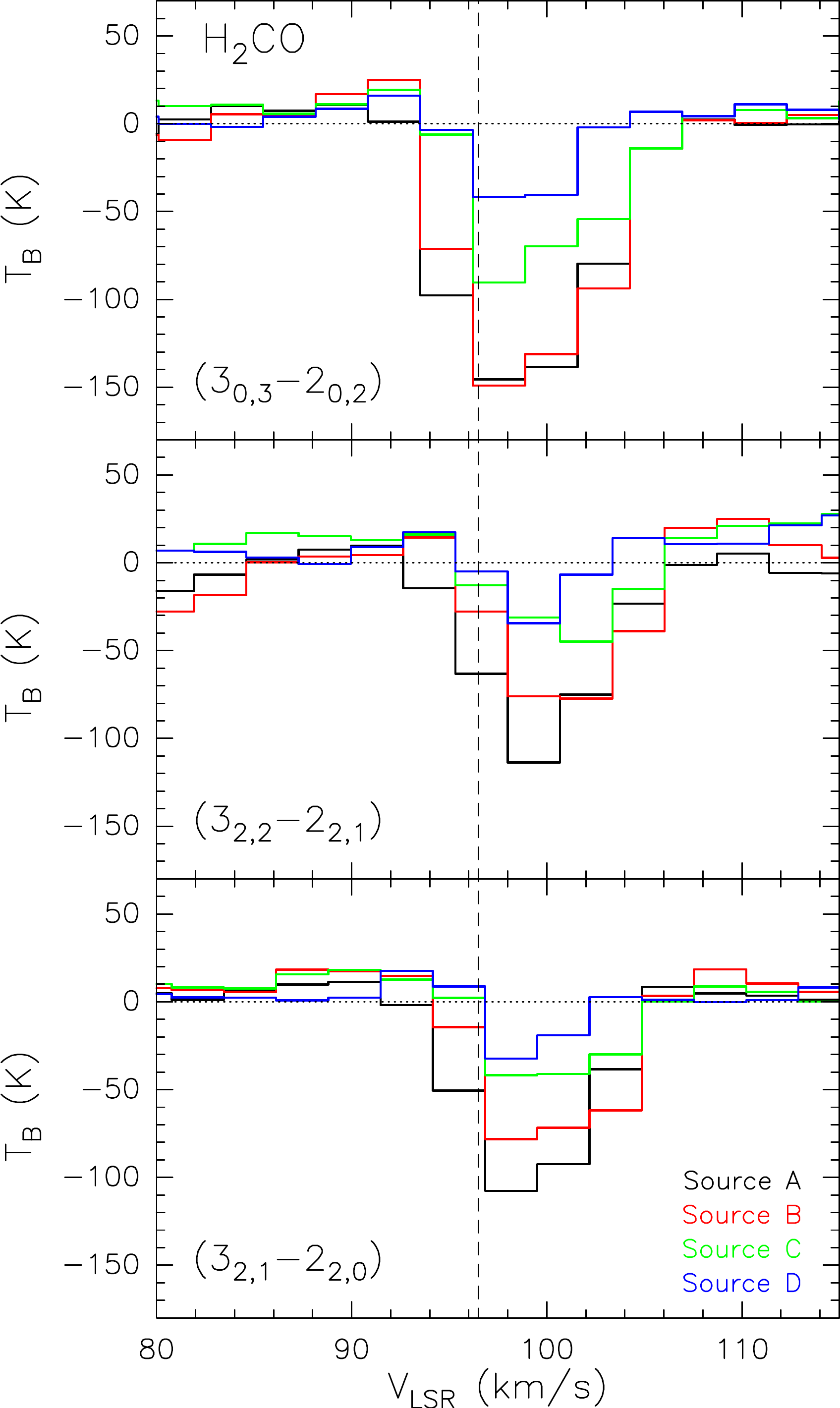}}
\caption{Same as Fig.~\ref{fig-spw3} for H$_2$CO (3$_{0,3}$--2$_{0,2}$) 
($E_{\rm up}/k_{\rm B}$=21\,K), (3$_{2,2}$--2$_{2,1}$) ($E_{\rm up}/k_{\rm B}$=68\,K), and (3$_{2,1}$--2$_{2,0}$) ($E_{\rm up}/k_{\rm B}$=68\,K). The $x$-axis is in units of velocity. The vertical dashed line indicates the systemic LSR velocity of 96.5$\pm0.5$\,km\,s$^{-1}$ (Beltr\'an et al.~\cite{beltran18}).
}
\label{fig-h2co}
\end{figure}

The spectra of the different $K$-components of CH$_3$CN\ (Figs.~\ref{fig-spw3} and \ref{fig-k3}) and, especially, of the different transitions of H$_2$CO (Fig.~\ref{fig-h2co}), integrated over the beam toward the position of the dust continuum emission peak of each source, confirm that the absorption is clearly red-shifted for all sources.  The systemic velocity of the {\it Main} core estimated from methyl cyanide is 96.5$\pm0.5$\,km\,s$^{-1}$ (e.g., Beltr\'an et al.~\cite{beltran18}). If the red-shifted absorption is interpreted as infall, this suggests that the material is still collapsing toward the sources and, therefore, that the young stellar objects have not reached their final mass yet. The deepest absorption of $\sim$$-150$\,K is  observed in H$_2$CO\,(3$_{0,3}$--2$_{0,2}$) toward sources A and B, while the absorption is $\sim$$-90$\,K for source C and $\sim$$-40$\,K for source D. The peak brightness temperature of the dust continuum emission at 1.4\,mm estimated by Beltr\'an et al.~(\cite{beltran21}) is $150\pm15$\,K (source A), $156\pm16$\,K (source B), $102\pm10$\,K (source C), and $48\pm5$\,K (source D). Therefore, this indicates that the absorbed fraction, which is between 80--100\%, is very high for all sources.

The red-shifted absorption is also clearly seen in CH$_3$CN toward sources A, B, and C (Figs.~\ref{fig-spw3} and \ref{fig-k3}), but it is less evident toward source D, when averaged over the beam. However, when searching pixel by pixel and channel by channel around the position of the dust continuum emission peak of source D, the line appears in absorption at different velocities. The deepest absorption in CH$_3$CN\ is observed for sources A and B, and reaches $\sim$$-90$\,K for CH$_3$CN\ $K=0$, while the absorption is $\sim$$-45$\,K for source C. The absorbed fraction in this case would be $\sim$60\% for sources A and B, and  $\sim$50\% for source C. Figure~\ref{fig-spw3} shows that the deepest red-shifted absorption occurs at different frequencies (or velocities) for the different sources, especially for sources A and B. This discrepancy in the velocity of the absorption is not visible in H$_2$CO (Fig.~\ref{fig-h2co}) because the spectral resolution of the observations, $\sim$2.7\,km\,s$^{-1}$, is not high enough. In Fig.~\ref{fig-k3}, we plot the spectra of the  CH$_3$CN $K=3$ transition only, with the x-axis in units of velocity, to better show the velocity of the red-shifted absorption. The velocity of the absorption for source A is $\sim$100.5$\pm$0.3\,km\,s$^{-1}$,  while the velocity for sources B and C is almost 2\,km\,s$^{-1}$\ higher, 102.2$\pm$0.3 and 102.5$\pm$0.3\,km\,s$^{-1}$, respectively. For source D, we have estimated the velocity of the deepest absorption pixel by pixel and found that for the $K=3$ transition the value is 104.8$\pm$0.3\,km\,s$^{-1}$. These velocities are between $\sim$4 and $\sim$8\,km\,s$^{-1}$\ higher than the systemic velocity of the {\it Main} core, which is 96.5$\pm0.5$\,km\,s$^{-1}$ (e.g., Beltr\'an et al.~\cite{beltran18}). 
     
The spectra of CH$_3$CN\ for source B shows a clear inverse P-Cygni profile for all the $K$-components (Figs.~\ref{fig-spw3} and ~\ref{fig-k3}). On the other hand, the spectra of source A and, although less evident, of source C show a much broader absorption profile with a tail extending to red-shifted velocities for source A and blue-shifted velocities for source C. In addition, the spectra of source A show a secondary absorption dip at blue-shifted velocities, at $\sim$$-$3\,km\,s$^{-1}$\, from the systemic velocity (see Figs.~\ref{fig-spw3} and \ref{fig-k3}). The shape of these profiles suggests the presence of different components that could either be related to the gas surrounding each source or indicate the presence of unresolved sources associated with cores A and C. The latter hypothesis would suggest that fragmentation has happened on scales smaller than the spatial resolution of our observations ($\sim$340\,au). Beltr\'an et al.~(\cite{beltran21}) have carried out 7\,mm observations at $\sim$2 times higher angular resolution ($\sim$0$\farcs05$ or $\sim$190\,au) with the VLA but have not found evidence of secondary companions for sources A and B, at least at scales of a few hundred au. Therefore, this suggests that the presence of different components is related to the gas emission itself.

\begin{table*}
\caption[] {Infall parameters.}
\label{table-infall}
\begin{tabular}{lcccccc}
\hline
&\multicolumn{1}{c}{$R_{\rm c}^a$} &
\multicolumn{1}{c}{$\langle n_{\rm H_2}\rangle^a$} &
\multicolumn{1}{c}{$V_{\rm inf}^{\rm H_2CO^b}$} &
\multicolumn{1}{c}{$V_{\rm inf}^{\rm CH_3CN^c}$} & 
\multicolumn{1}{c}{$\dot M_{\rm inf}^{d, e}$} &
\multicolumn{1}{c}{$\dot M_{\rm acc}^{f}$}
\\
\multicolumn{1}{c}{Source} &
\multicolumn{1}{c}{(au)} &
\multicolumn{1}{c}{(10$^9$\,cm$^{-3}$)} &
\multicolumn{1}{c}{(km\,s$^{-1}$)} &
\multicolumn{1}{c}{(km\,s$^{-1}$)}  &
\multicolumn{1}{c}{(10$^{-2}\,M_\sun$\,yr$^{-1}$)} &
\multicolumn{1}{c}{(10$^{-4}\,M_\sun$\,yr$^{-1}$)}\\
\hline
A  &414$\pm$19 &7.0$\pm$1.2 &1.9$\pm$2.7 &4.0$\pm$0.33  &2.4$\pm$3.4 & 0.04--2.5\\ 
B  &390$\pm$22 &7.6$\pm$1.3 &1.9$\pm$2.7 &3.7$\pm$0.33 &2.1$\pm$3.0 & 0.014--0.9\\
C  &535$\pm$30 &5.2$\pm$1.1 &2.8$\pm$2.7 &6.0$\pm$0.33 &4.4$\pm$4.3 & 0.003--0.2\\
D  &456$\pm$41 &8.2$\pm$2.5 &1.9$\pm$2.7 &8.3$\pm$0.33 &3.4$\pm$4.9 & 0.04--2.7\\
\hline
\end{tabular}
\\
$^a$ Estimated from dust continuum observations at 3.5\,mm and $\sim$0$\farcs075$ resolution by Beltr\'an et al.~(\cite{beltran21}).   \\  
$^b$ Mean value of the three transitions. \\ 
$^c$ Calculated from CH$_3$CN\ $K=3$. \\
$^d$ Calculated using  $V_{\rm inf}^{\rm H_2CO}$. \\
$^e$ For infall confined to a solid angle $\Omega$, the infall rate scales linearly with $\Omega/(4\,\pi$). \\
$^f$ Estimated from the mass outflow rate assuming a gas temperature of 50\,K,  an SiO abundance range of  10$^{-8}$--10$^{-7}$ (see Table~\ref{table-outflow}), and a fraction of accreting gas that is launched in the jet in the  range 0.1--0.3. 
\end{table*}

\subsubsection{Infall in individual sources}

From the red-shifted absorption it is possible to estimate the infall rate, $\dot M_{\rm inf}$, if one knows the density and the size of the core, and assuming that the infall velocity, $V_{\rm inf}$,  is equal to the difference between the velocity of the absorption feature and the systemic LSR velocity. Assuming that the material is free falling, and therefore, that  $V_{\rm inf}$ is proportional to $R^{-1/2}$, the infall rate inside a solid angle $\Omega$ can be estimated from this expression (see Appendix A for the detailed derivation of the mass infall rate):
\begin{equation}
\dot M_{\rm inf} = \frac{\Omega}{2}\, V_{\rm inf}(R_{\rm c})\, \mu\, m_{\rm H}\langle n_{\rm H_2}\rangle R_{\rm c}^2,
\end{equation}
where $R_{\rm c}$ is the radius of the core, $V_{\rm inf}(R_{\rm c})$ is the infall velocity at $R_{\rm c}$, $m_{\rm H}$ the mass of the H atom, $\mu=2.8$ the mean molecular weight, and $\langle n_{\rm H_2}\rangle$ the mean number density in the core. Both $R_{\rm c}$ and $\langle n_{\rm H_2}\rangle$ have been estimated from dust continuum observations at 3.5\,mm and $\sim$0$\farcs075$ resolution by Beltr\'an et al.~(\cite{beltran21}). Although $V_{\rm inf}$ can be easily obtained from the spectra, the radius at which this infall velocity is associated is not known. Based on fact that the upper level energies of the observed transitions of H$_2$CO are lower than those of the CH$_3$CN\ ones, we have assumed that the former species is tracing cooler material located closer to the surface of the core than the latter. Therefore, to estimate $\dot M_{\rm inf}$, we have assumed that  $V_{\rm inf}(R_{\rm c})$ is similar to the infall velocity estimated from H$_2$CO.   Because  the velocity of the absorption is different depending on the H$_2$CO transition, as seen in Fig.~\ref{fig-h2co}, to estimate $V_{\rm inf}(R_{\rm c})$ and, therefore, $\dot M_{\rm inf}$, we have used the mean of the infall velocity obtained from each of the H$_2$CO transitions.

Table~\ref{table-infall} shows the $\dot M_{\rm inf}$ estimated for each source embedded in the {\it Main} core together with $V_{\rm inf}$ estimated from H$_2$CO. The infall rates have been estimated assuming $\Omega=4\,\pi$, because the fact that infall is observed in all the sources suggests that it cannot be very radially asymmetric, otherwise it would be statistically unlikely to observe collapse in all of them, unless they are aligned. Note that these infall rates should be considered as upper limits, because $\Omega$ could be smaller than $4\,\pi$ and the infall velocity at $R_{\rm c}$ could be even smaller than that obtained from H$_2$CO. Table~\ref{table-infall} also shows, for comparison, the infall velocities obtained from CH$_3$CN\ $K=3$. As expected, because CH$_3$CN\ is probably tracing the embedded dense gas deeper than H$_2$CO, 
these infall velocities are greater. Note that the infall rates estimated using the infall velocities obtained from CH$_3$CN\ $K=3$ would be a factor $\sim$2--4 higher than those estimated with H$_2$CO.

As seen in this table, the infall rates are high and similar for all the sources and  are similar to the infall rate of 0.028\,$M_\odot$\,yr$^{-1}$ estimated for the whole {\it Main} core by Beltr\'an et al.~(\cite{beltran21}), assuming a mean density of $\sim$4$\times10^7$\,cm$^{-3}$ and a radius of $\sim$4000\,au. Red-shifted H$_2$$^{18}$O\,(1$_{11}$--0$_{00}$) absorption observed with ${\it Herschel}$ by van der Tak et al.~(\cite{vandertak19}) indicates a similar infall rate of $\sim$10$^{-2}$\,$M_\odot$\,yr$^{-1}$ in G31.

Infall rates on the order of $\sim$10$^{-2}$\,$M_\odot$\,yr$^{-1}$ for all four sources imply that the mass of the {\it Main} core, $\sim$70\,$M_\odot$ (Cesaroni et al.~\cite{cesa19}), would be incorporated in the sources in $\sim$1000\,yr. This accretion timescale of the {\it Main} core is similar to the free-fall times of the individual sources A to D, which are $\sim$700--900\,yr.  This suggests that the sources are not dynamically individual objects but evolve together with the {\it Main} core in which they are embedded, i.e., the stars inside are not forming from the individual mass reservoirs of sources A to D, but have the entire mass reservoir of the {\it Main} source available to build up more massive stars. The accretion timescale of the {\it Main} core is an order of magnitude smaller than its rotation timescale, which is $\sim$4$\times10^4$\,yr assuming a rotation velocity of $\sim$3\,km\,s$^{-1}$\ (estimated as half the range of the velocity gradient) and a radius of $\sim$4000\,au (see Fig.~\ref{fig-moments}). This confirms that rotating toroids, which is what the HMC G31 core is (Beltr\'an et al.~\cite{beltran05}), are non-equilibrium, transient collapsing structures that must be constantly replenished with fresh material from a large-scale material reservoir (e.g., Kuiper \& Hosokawa~\cite{kuiper18}). For G31, this reservoir of material has recently been identified by means of N$_2$H$^+$ observations of the parsec-scale IR-dark clump surrounding the HMC (Beltr\'an et al., in preparation).

\begin{figure*}
\centerline{\includegraphics[angle=0,width=14.6cm]{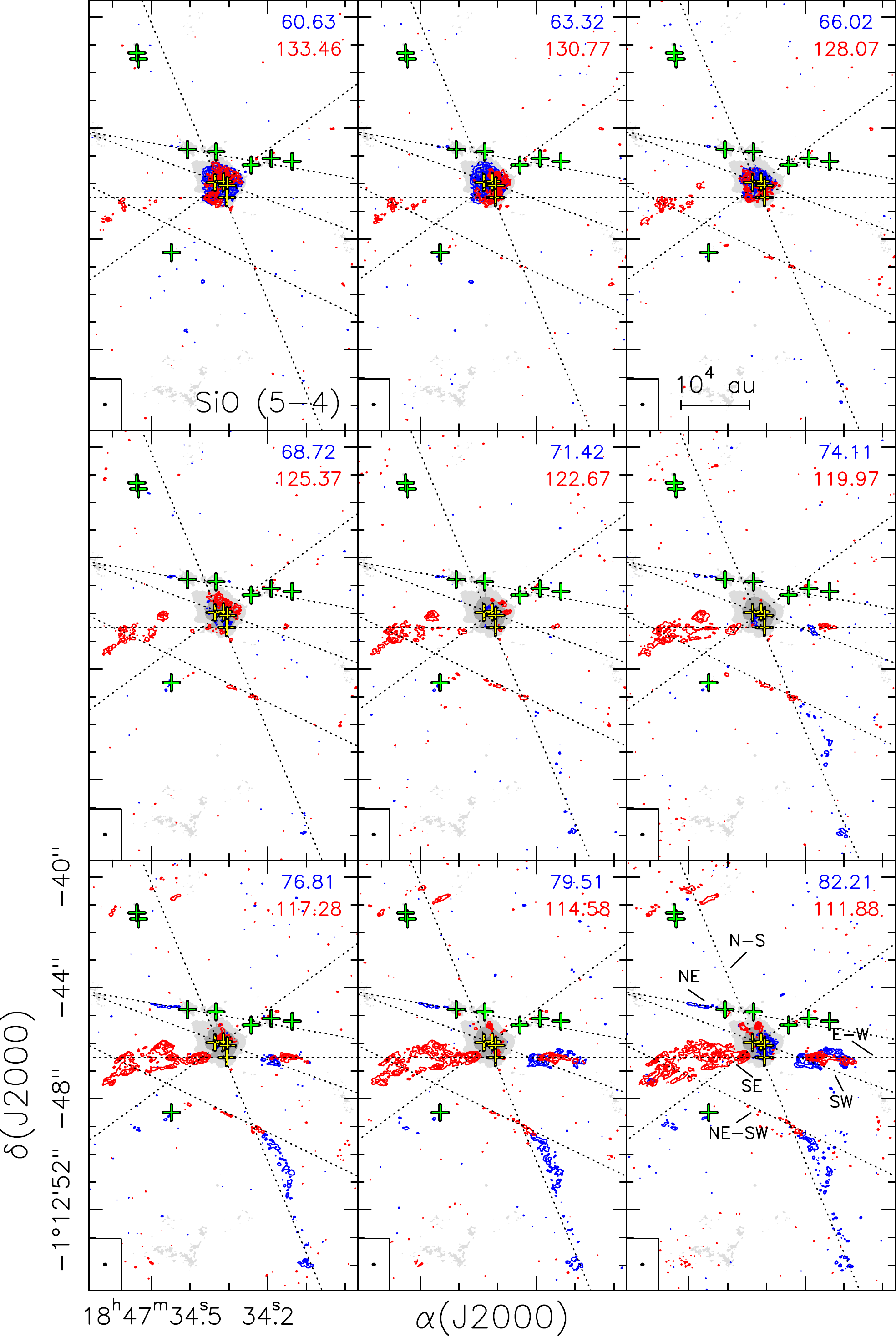}}
\caption{Velocity channel maps of the SiO\,(5--4) line emission ({\it colors}) overlaid on the 1.4\,mm continuum emission ({\it grayscale}). Each box contains pairs of maps corresponding to the blue- ({\it blue contours}) and red-shifted ({\it red contours}) emission at about the same velocity offset (in absolute value) from the the systemic velocity (96.5$\pm0.5$\,km\,s$^{-1}$). The corresponding LSR velocities are indicated in the top right of the box. The conversion factor from Jy\,beam$^{-1}$ to K, in the Rayleigh-Jeans approximation, is 2844. Contour levels are 3, 6, 9, 12, 15, and 20 times 0.55\,mJy/beam$^{-1}$. Grayscale levels are 3, 20, 83, and 165 times 1\,$\sigma$ times 0.32\,mJy/beam$^{-1}$. The synthesized beam of $0\farcs11\times0\farcs08$, P.A.\ =$-83^\circ$ is shown in the lower left-hand corner of each panel. The black dotted lines indicate the direction of the six outflows identified in G31 in this work. The labels of the outflows are indicated in the bottom right panel. The yellow crosses indicate the positions of the four continuum sources embedded in the {\it Main} core and the green crosses mark the positions of the rest of continuum sources detected in the region (Beltr\'an et al.~\cite{beltran21}).}
\label{fig-sio-chan1}
\end{figure*}

\begin{figure*}
\addtocounter{figure}{-1}
\centerline{\includegraphics[angle=0,width=14.6cm]{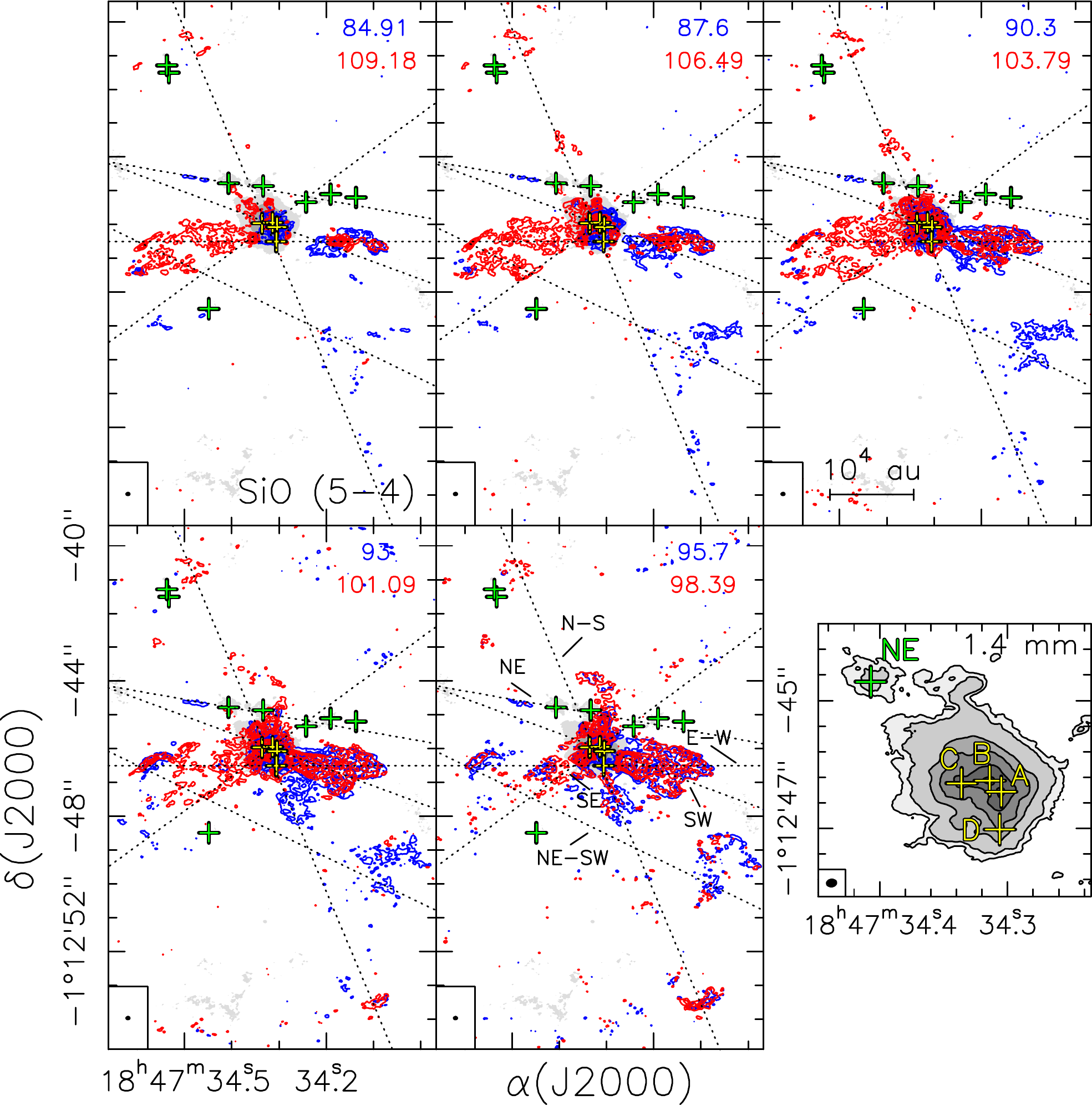}}
\caption{Continued. Colored contour levels are 3, 6, 9, 12, 15, 20, and 30 times 0.7\,mJy/beam$^{-1}$. The bottom right panel is a close-up of the central region to better show the 1.4\,mm continuum emission and the position of sources A, B, C, D, and NE. Grayscale and contour levels are 3, 6, 24, 48, 96, and 165 times 1\,$\sigma$ times 0.32\,mJy/beam$^{-1}$.}
\label{fig-sio-chan2}
\end{figure*}

\subsection{Outflow emission}
\label{sect-sio}

Observations of CO, $^{13}$CO, and SiO have revealed the presence of several molecular outflows in G31 (Olmi et al.~\cite{olmi96}; Cesaroni et al.~\cite{cesa11}; Beltr\'an et al.~\cite{beltran18}). The SiO emission observed at $\sim$0$\farcs2$ ($\sim$750\,au) resolution by Beltr\'an et al.~(\cite{beltran18}) has revealed the presence of at least three molecular outflows in the region: a strong east-west (E--W) outflow, and two weaker north-south (N--S) and northeast-southwest (NE--SW) outflows. 

The new SiO\,(5--4) ALMA observations at $\sim$0$\farcs09$ ($\sim$340\,au) 
reveal the presence of additional molecular outflows in the G31 HMC as shown in Fig.~\ref{fig-sio-chan1}. In each box of this figure, we show the SiO channel maps corresponding to the blue-shifted and red-shifted emission at about the same velocity offset (in absolute value) from the systemic velocity (96.5$\pm0.5$\,km\,s$^{-1}$). The blue-shifted and red-shifted emission observed at high velocities (from $\sim$ 60 to 69\,km\,s$^{-1}$\ for the blue-shifted emission and from 125 to 139\,km\,s$^{-1}$ for the red-shifted emission) toward the HMC is probably contaminated by  emission of a high-density tracer. Possible contaminants
are ethylene glycol, (CH$_2$OH)$_2$, at blue-shifted velocities and cyanide radical, $^{13}$CN, at red-shifted velocities. Both species have been previously detected in G31 (Rivilla et al.~\cite{rivilla17}; Mininni, private communication).

\begin{figure*}
\centerline{\includegraphics[angle=0,width=17cm]{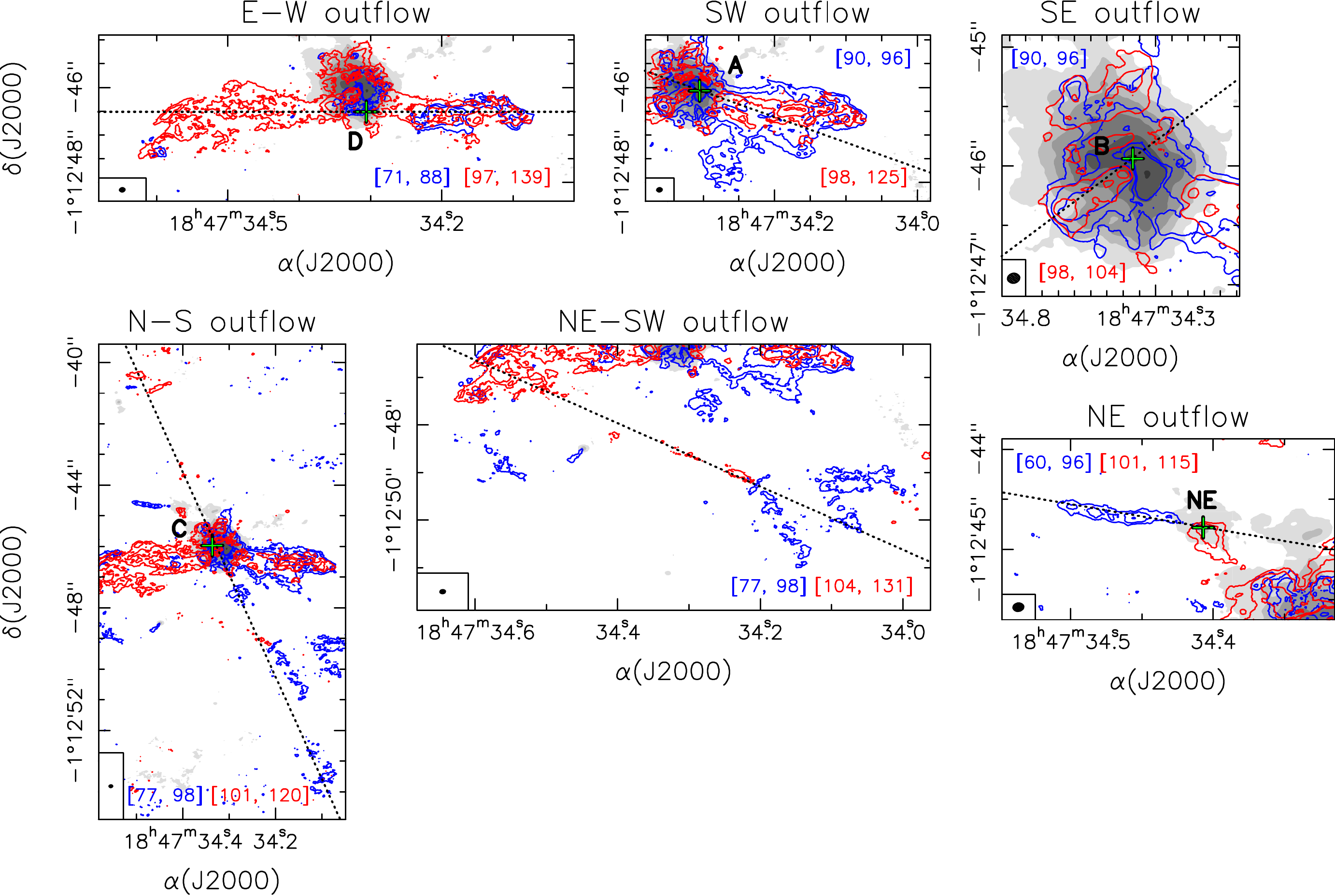}}
\caption{Overlay of the ALMA 1.4\,mm continuum emission ({\it grayscale}) on the blue-shifted ({\it blue contours}) and red-shifted ({\it red contours}) SiO\,(5-- 4) emission averaged over different velocity intervals for the six outflows detected in the region. The velocity ranges within which the blue-shifted and red-shifted emission were averaged is indicated within brackets in each panel.  The systemic LSR velocity of G31 is 96.5\,km\,s$^{-1}$. Contour levels are 3, 6, 12, and 24 times 0.35\,mJy/beam$^{-1}$ ({\it blue} and {\it red}) for the E--W and NE--SW outflows, 0.6\,mJy/beam$^{-1}$  ({\it blue}) and 0.4\,mJy/beam$^{-1}$  ({\it red}) for the SW outflow, 1.0\,mJy/beam$^{-1}$  ({\it blue}) and 1.5\,mJy/beam$^{-1}$  ({\it red}) for the SE outflow, 0.36\,mJy/beam$^{-1}$  ({\it blue}) and 0.4\,mJy/beam$^{-1}$  ({\it red}) for the N--S outflow, and 0.22\,mJy/beam$^{-1}$  ({\it blue}) and 0.32\,mJy/beam$^{-1}$  ({\it red}) for the NE outflow. The conversion factor from Jy\,beam$^{-1}$ to K is 2844. Grayscale contours for the continuum emission are 0.96, 3.2, 6.4, 14.4, 28.8, and  54.4\,mJy/beam$^{-1}$. The black dotted line indicates the direction of the outflow and the green cross the position of the suggested driving source.}
\label{fig-six-outflows}
\end{figure*}

\begin{table*}
\caption[] {Outflow properties calculated from SiO.}
\label{table-outflow}
\tabcolsep 3pt 
\begin{tabular}{lcccccc}
\hline
\multicolumn{1}{c}{Properties} &
\multicolumn{1}{c}{SW} &
\multicolumn{1}{c}{SE} &
\multicolumn{1}{c}{N--S} &
\multicolumn{1}{c}{E--W} &
\multicolumn{1}{c}{NE} &
\multicolumn{1}{c}{NE--SW} 
\\
\hline
Driving source  &A &B &C &D &NE &? \\
$M_{\rm core}^{a}\,(M_\odot)$ &16$\pm$3 &15$\pm$3 &26$\pm$5 &26$\pm$8 &18--38  & $\lesssim 0.3^{b}$ \\
$M_{\rm out}\,(M_\odot)$   & 0.05--0.61  &0.015--0.17  &(0.9--11)$\times10^{-2}$ &0.05--0.57 &(0.3--3)$\times10^{-2}$ &0.03--0.35 \\
$P_{\rm out}\,(M_\odot$\,km\,s$^{-1}$)                           &0.26--2.9 &0.07--0.82 &0.1--1.1 &0.65--7.3 &0.03--0.35 &0.35--3.9 \\
$E_{\rm out}\,(L_\odot\,$yr)                               &(0.17--1.9)$\times10^3$   &40--456 & (0.94--10.7)$\times10^3$ &(0.13--1.4)$\times10^3$ &43--494  &(0.49--5.5)$\times10^3$ \\
$\dot M_{\rm out}\,(10^{-4}\, M_\odot\,$yr$^{-1})$            &0.23--2.5 &0.08--0.9 &0.02--0.2 &0.24--2.7 &0.04--0.4 &0.09--1.1 \\
$\dot P_{\rm out}\,(10^{-3} M_\odot$\,km\,s$^{-1}$\,yr$^{-1})$   &0.11--1.2 &0.04--0.46 &0.02--0.18 &0.31--3.5 &0.04--0.44 &0.11--1.2 \\
$\dot E_{\rm out}\,(L_\odot)$   &0.07--0.79 &0.02--0.25 &0.02--0.23 &0.45--5.1 &0.05--0.62 &0.15--1.7     \\
$\langle R_{\rm lobe}\rangle$\,(pc) &0.07 &0.02 &0.15 &0.07 &0.03 &0.09 \\ 
$V_{\rm max}$ (km\,s$^{-1}$) &28.5 &10 &24 &34 &36 &28 \\
$t_{\rm dyn}$\,($10^3$\,yr)  &2.4 &1.8 &6.0 &2.1 &0.8 &3.2  \\
\hline
\end{tabular}
\\
Parameters estimated assuming a range of gas temperature of 30--50\,K, and a range of abundance relative to H$_2$ of $10^{-8}$--$10^{-7}$. The derived parameters should be taken as lower limits since the interferometer filters out extended emission that most likely is associated with these outflows. \\
$^a$ Mass of the core. Estimated from dust continuum observations at 3.5\,mm and $\sim$0$\farcs075$ resolution by Beltr\'an et al.~(\cite{beltran21}).  
$^b$ Upper limit estimated from dust continuum observations at 1.4\,mm assuming a dust temperature of 20\,K. See Sect.~\ref{sect-indi-out}\\  
\end{table*}

\begin{figure*}
\centerline{\includegraphics[angle=0,width=17cm,angle=0]{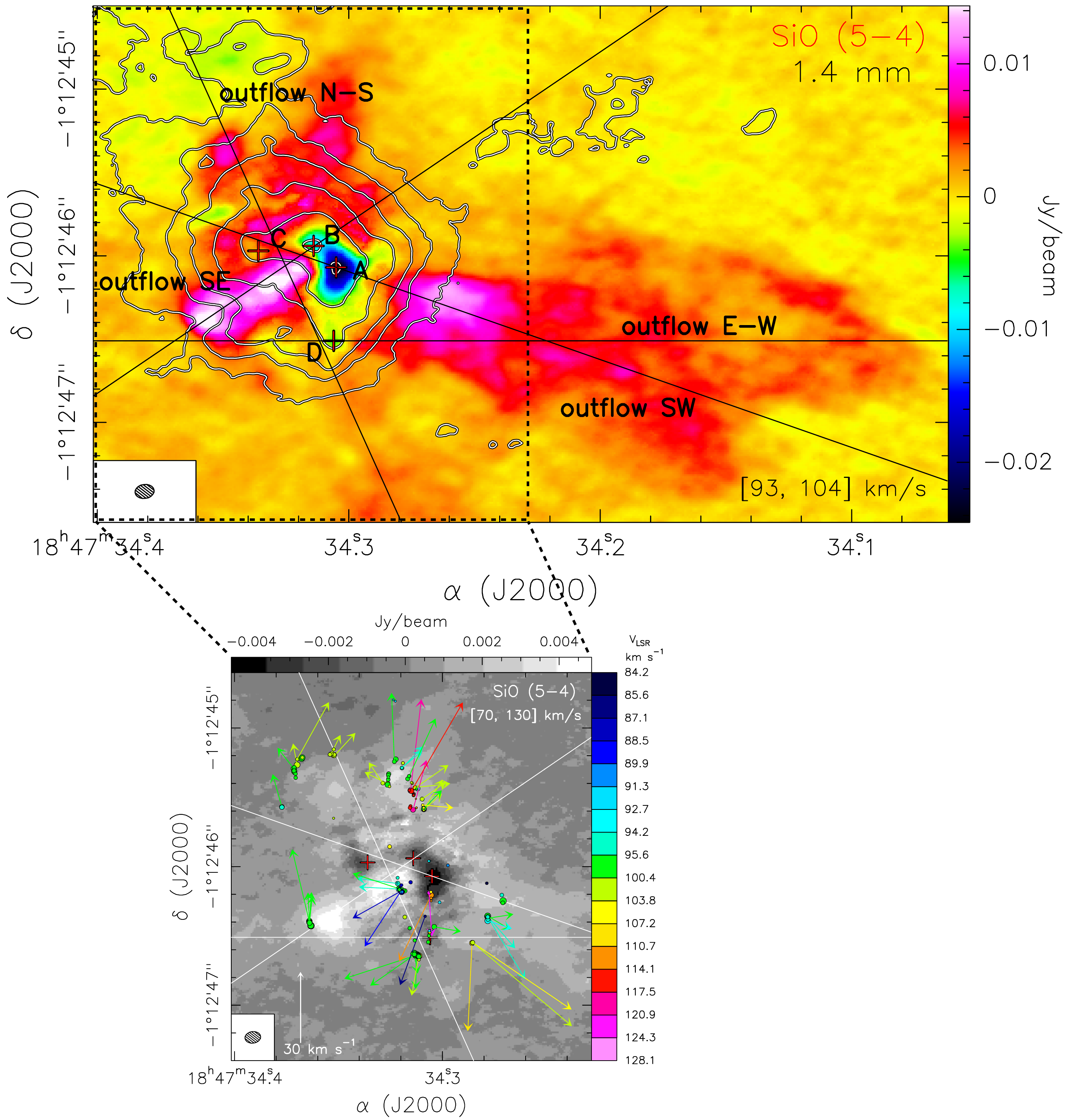}}
\caption{{\it Top panel}: Overlay of the ALMA 1.4\,mm continuum emission ({\it grayscale}) on the SiO\,(5--4) emission averaged over the (93, 104)\,km\,s$^{-1}$\ velocity interval. Contour levels are the same as in Fig.~\ref{fig-sio-chan1}. {\it Bottom panel}: Close-up of the central region. Colored circles mark the position of H$_2$O masers while colored vectors indicate the direction and the amplitude of the proper motions (Moscadelli et al.~\cite{mosca13}). The white vector in the bottom left corner indicates the amplitude scale of proper motions in kilometer per second. The synthesized beam is drawn in the bottom left corner of each panel. The red crosses mark the positions of the continuum sources embedded in the {\it Main} core (Beltr\'an et al.~\cite{beltran21}). The black (white) lines indicate the direction of the outflows driven by sources A, B, C, and D.}
\label{fig-sio-zoom-A}
\end{figure*}

\subsubsection{Individual outflows}
\label{sect-indi-out}

The SiO emission suggests the presence of at least six outflows in the G31 star-forming region, which are highlighted in Fig.~\ref{fig-six-outflows}, where we have plotted the emission averaged over different velocity intervals. We describe below each of the outflows and discuss their possible driving sources. Note that because of the complexity of the region, with several sources and outflows within few 1000s au, the attribution of a certain driving source to a given outflow is further complicated by possible changes in the outflow direction.

\begin{itemize}

\item {\it E--W outflow:} The E--W outflow, which is the strongest in the region, is clearly detected at all velocity intervals (at least the red-shifted emission). Note that, as suggested by Beltr\'an et al.~(\cite{beltran18}), the eastern red-shifted lobe could be contaminated by emission from the NE--SW outflow, especially at low velocities. Source D could be driving this outflow, as suggested by Beltr\'an et al.~(\cite{beltran21}), because it is located at about the center of symmetry of the eastern and western lobes. 

\item {\it SW outflow:} The velocity channel maps (Fig.~\ref{fig-sio-chan1}) suggest that part of the SiO emission of the western lobe of the E--W outflow could be powered by source A. The presence of SiO emission that appears to be driven by source A starts to be visible for velocities higher than $\sim$90\,km\,s$^{-1}$ and is more evident for channels with velocities closer to the systemic LSR velocity of $\sim$96.5\,km\,s$^{-1}$, as seen in Fig.~\ref{fig-sio-zoom-A}, where we plot the SiO emission averaged over the (93, 104)\,km\,s$^{-1}$\ velocity interval overlaid on the dust continuum emission at 1.4\,mm. The SiO emission westward of the HMC shows two molecular lobes that end in bow shocks, with the shorter lobe clearly pointing to the position of source A. The longer lobe, which shows a clear bow shock shape, could be associated with the E--W outflow or, alternatively, with source A. From the observations, it is not possible to distinguish between these two possibilities. Note that toward the position of the dust continuum sources, SiO is seen in absorption, making it difficult to clearly associate the outflow emission with its corresponding powering source. Summarizing, the SiO observations suggest the presence of two overlapping outflows to the southwest of the {\it Main} core: the E--W outflow, driven by source D, with red-shifted emission mainly eastward of source D and blue-shifted emission, with velocities between $\sim$71 and $\sim$90\,km\,s$^{-1}$, to the west; and the SW outflow, driven by source A, with a clear southwestern lobe of overlapping blue-shifted and red-shifted emission. Note that the red-shifted emission is possibly contaminated by some  red-shifted emission from the E--W outflow.

\item {\it SE outflow:}  Figure~\ref{fig-sio-zoom-A} points to the presence of an outflow associated with source B. As seen in this figure and also in the channel maps of Fig.~\ref{fig-sio-chan1}, the strongest SiO emission seems to delineate a collimated molecular outflow from source B, that we name outflow SE. This outflow is better traced at low blue-shifted and red-shifted velocities with respect to the systemic velocity, in particular for the velocity interval (90, 103.8)\,km\,s$^{-1}$.

\item {\it N--S outflow:} The direction of the N--S outflow is mainly traced by the southernmost and northernmost knots, which are visible from low to intermediate velocities. The  southernmost knot shows a clear bow-shock shape at velocities closer to the systemic velocities (Fig.~\ref{fig-sio-chan1}).  Regarding its powering source, Beltr\'an et al.~(\cite{beltran18}) suggested that it could be one of the two free-free sources detected by Cesaroni et al.~(\cite{cesa10}), which Beltr\'an et al.~(\cite{beltran21}) identified as sources A and B. As seen in Figs.~\ref{fig-sio-chan1} and \ref{fig-six-outflows}, the line connecting the red-shifted and blue-shifted knots crosses the position of source D and passes slightly west of source C. Taking into account the poor accuracy of the direction of the outflow and the fact that source D seems to be driving the E--W outflow, source C appears to be the best candidate to power the N--S outflow. 

\item {\it NE--SW outflow:} The NE--SW SiO molecular outflow proposed by Beltr\'an et al.~(\cite{beltran18}) and also visible in the CO channel maps of Cesaroni et al.~(\cite{cesa11}), which is located southward of the {\it Main} core, is visible at all velocity intervals, especially the red-shifted emission. However, Beltr\'an et al.~(\cite{beltran21}) have not detected any dust continuum emission source at a 3\,$\sigma$ level of $\sim$1\,mJy\,beam$^{-1}$\ that could be a candidate to drive this outflow.  Assuming a dust opacity of 1.0\,cm$^2$\,g$^{-1}$ at 1.4\,mm (Ossenkopf \& Henning~(\cite{ossenkopf94}) for a Mathis, Rumpl, \& Nordsieck~(\cite{mathis77}; MRN) distribution with thin ice mantles and a gas density of 10$^8$\,cm$^{-3}$), a mass-to-dust ratio of 100, and a dust temperature of 20\,K, the mass of the driving source would be $\lesssim 0.3$\,$M_\odot$. This suggests that the source powering this outflow is a low-mass young stellar object. Note that if the dust temperature were $>20$\,K, then the mass would be even smaller.

\item {\it NE outflow:} The channel maps (Fig.~\ref{fig-sio-chan1}) and, in particular, the averaged emission maps (Fig.~\ref{fig-six-outflows}) indicate the presence of an additional molecular outflow that could be associated with the {\it NE} core. This SiO outflow, named outflow NE, shows a very collimated blue-shifted lobe with emission from low- to high velocities. On the other hand, the red-shifted emission is hardly visible very close to the position of the source. 

\end{itemize}

Summarizing, the high-angular resolution SiO emission observations have resolved for the first time the complicated outflow emission in the {\it Main} core of G31 and have suggested that each of the four dust continuum emission sources embedded in it could be driving a molecular outflow. This confirms what has already been suggested by the red-shifted absorption observed in CH$_3$CN\ and H$_2$CO, that is, that the sources are still actively accreting and therefore, gaining mass. Furthermore, these observations confirm that also in the high-mass regime, accretion appears to be commonly associated with outflow activity. The bottom panel of Fig.~\ref{fig-sio-zoom-A} shows a close-up view of the SiO averaged emission toward the {\it Main} core overlaid with the H$_2$O maser positions and proper motions from Moscadelli et al.~(\cite{mosca13}). The proper motions confirm the existence of expansion motions in  the core. In fact, except for a group of masers located northward of source B and a small  group located at the southern edge of the SE outflow, the water masers positions and their proper motions are clearly associated with the four SiO outflows in the core: outflows SE, E--W, N--S, and SW.

\subsubsection{Physical properties of the outflows}

The physical properties of the SiO outflows have been estimated by assuming a range of temperatures of 30 and 50\,K, following Beltr\'an et al.~(\cite{beltran18}). The lower limit is set by the SiO peak brightness temperature. For the  abundance of SiO relative to H$_2$ we used a range of 10$^{-8}$--10$^{-7}$ (e.g., Codella et al.~(\cite{codella13}).  Table~\ref{table-outflow} shows the mass, $M_{\rm out}$, of the six outflows detected in the region, their momentum, $P_{\rm out}$, their kinetic energy, $E_{\rm out}$, and their corresponding rates obtained by dividing the previous quantities by the dynamical timescale of each outflow, $t_{\rm dyn}$, average length of the red-shifted and blue-shifted lobes, $\langle R_{\rm lobe}\rangle$, the average red-shifted and blue-shifted maximum velocity of the outflow, $V_{\max}$, and  $t_{\rm dyn}$. The dynamical timescales have been calculated by dividing $\langle R_{\rm lobe}\rangle$ by $V_{\max}$, and as seen in the table, are quite similar and of the order of $10^3$\,yr. Note that the outflow properties have not been corrected for the (unknown) inclination angle, $i$, of the flows with respect to the plane of the sky. In case of correcting for inclination, $V_{\max}$ should be divided by $\sin i$, $\langle R_{\rm lobe}\rangle$ by $\cos i$, and $t_{\rm dyn}$ should be multiplied by $\tan i$.

The outflows with the highest mass loss rate are outflow E--W, associated with source D, and outflow SW, driven by source A, while the weakest one is outflow N--S, driven by source C. Outflow SE is the one with the smallest velocities and the smallest lobes. However, $V_{\rm max}$ and $\langle R_{\rm lobe}\rangle$ should be considered as lower limits because the emission of the red-shifted lobe is highly blended with that of the red-shifted lobe of outflow E--W. 

We have found no correlations between the properties of the outflows $\dot M_{\rm out}$ and $M_{\rm out}$ and the infall parameters $\dot M_{\rm inf}$ and $V_{\rm inf}^{\rm CH_3CN}$, nor with the mass of the sources. This is not surprising taking into account the missing flux of the extended SiO emission, the low statistics (only 4 sources), and the uncertainties in the derivation of the parameters, such as the SiO abundance which can vary by a factor 10 or the radius associated with the infall velocity, which is needed to estimate $\dot M_{\rm inf}$.

\begin{figure*}
\centerline{\includegraphics[angle=0,width=17cm]{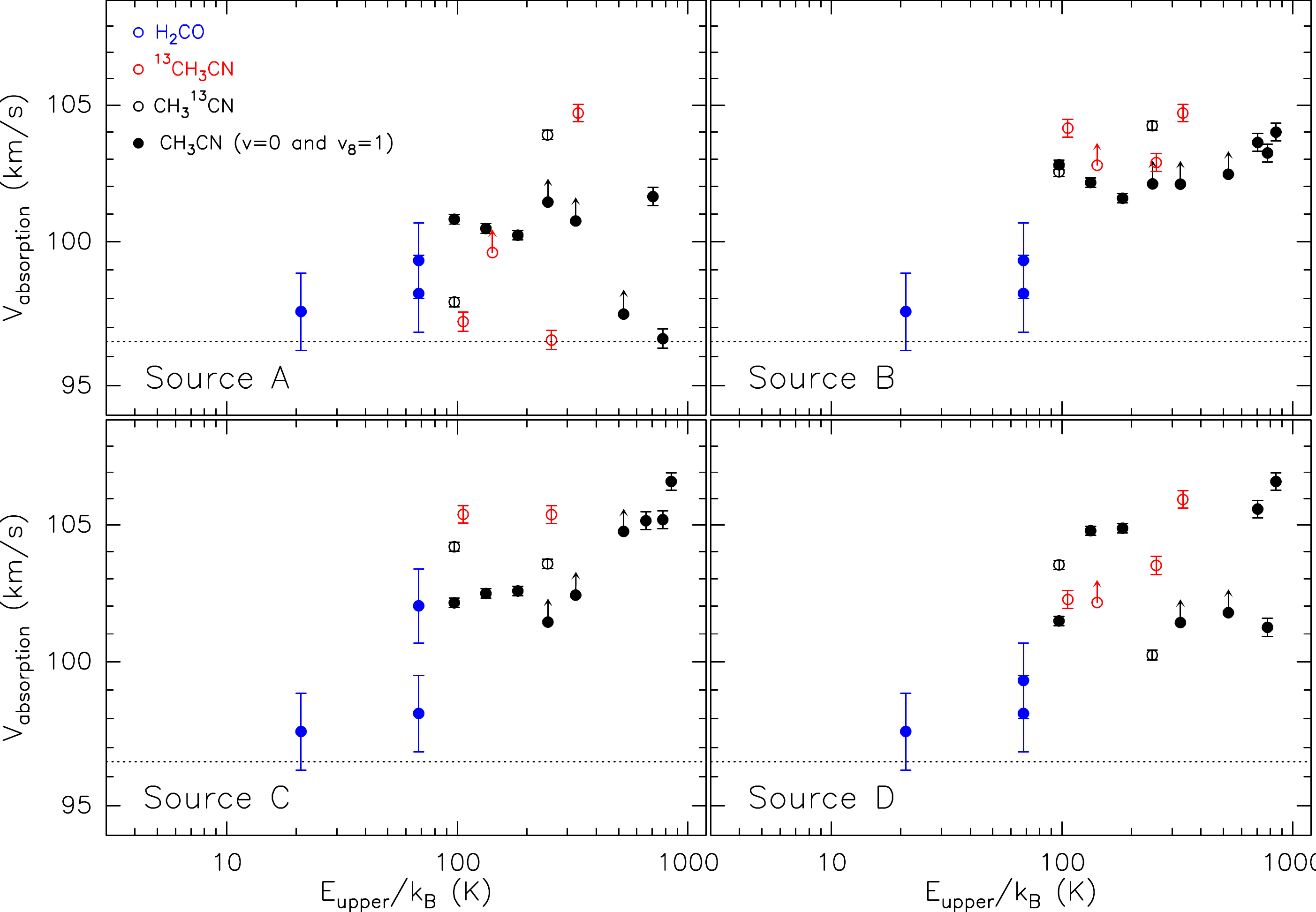}}
\caption{Velocity of the absorption feature, measured toward the dust continuum peak of each source embedded in the {\it Main} core of G31 versus the excitation energy of the corresponding CH$_3$CN and CH$_3^{13}$CN\,(12--11), $^{13}$CH$_3$CN\,(13--12), and H$_2$CO\,(3--2) lines. The systemic LSR velocity is 96.5\,km\,s$^{-1}$\ and is indicated with a dotted line. The black and red  arrows indicate lower limits of the velocity and are for those lines that are slightly blended with other lines at frequencies lower than their rest frequency.}
\label{fig-infall}
\end{figure*}

\section{Discussion}

\subsection{Accelerating infall}

Beltr\'an et al.~(\cite{beltran18}) concluded that the infall was accelerating toward the center of the {\it Main} core by measuring the velocity of the absorption feature for different H$_2$CO, CH$_3$CN and isotopologues transitions with different excitation energies and observing that the infall velocity increases with the line excitation energy. With the higher angular resolution of  the observations presented here we can investigate whether the infall is accelerating within each source. The red-shifted velocity has only been calculated for those lines that are not heavily blended with other species, because blending affects the estimate. For a few transitions that are slightly blended, we estimated a lower limit for the velocity. In Fig.~\ref{fig-infall}, we plot the red-shifted velocity as a function of the excitation energy of the CH$_3$CN (ground state and vibrationally excited), CH$_3^{13}$CN, $^{13}$CH$_3$CN, and H$_2$CO lines. As seen in this figure, for all but source A, the velocity of absorption is lower for the H$_2$CO transitions, which have the lowest excitation energies. This suggests that H$_2$CO is probably tracing the infall at the outer regions of the sources. Taking into account that the velocity of the absorption feature seen in the different H$_2$CO transitions is similar for all the sources, these outer regions could be indeed located at the common envelope/core surrounding the sources.  Figure~\ref{fig-infall} shows that, although with some dispersion, the red-shifted velocity for sources B, C, and D increases with the upper level energy of the transition, suggesting that infall could be accelerating also within the sources. Regarding source A, as mentioned in Sect.~\ref{sect-ch3cn}, the spectra of CH$_3$CN clearly indicate the presence of different components at blue-shifted and red-shifted velocities, and therefore, it is difficult to properly estimate the velocity of the red-shifted absorption. This is evident in Fig.~\ref{fig-infall}, where one can see that the dispersion of the estimates is too large to draw any conclusion about the infall.

\subsection{Infall versus accretion}

The four sources embedded in the {\it Main} core of G31 are associated simultaneously with infall and outflow. This is not surprising, as both mechanisms, ejection and collapse of material, are closely related to each other in the star-formation process. The SiO observations have allowed us to derive the outflow rate, $\dot M_{\rm out}$, that, after taking into account a few assumptions, could allow us to estimate the mass accretion rate onto the central star, $\dot M_{\rm acc}$. Assuming conservation of the momentum rate between the outflow and the jet powering it,  $\dot M_{\rm out}$ is related to the mass loss rate of the 
internal jet $\dot M_{\rm jet}$ as $\dot M_{\rm out} =\dot M_{\rm jet}\,V_{\rm jet}/V_{\rm out}$. Assuming furthermore a ratio between the jet velocity, $V_{\rm jet}$, and the molecular outflow velocity, $V_{\rm out}$, we can estimate $\dot M_{\rm acc}$ if we know the fraction of accreting gas that is launched in the jet. Beuther et al.~(\cite{beuther02}) have estimated $V_{\rm jet}/V_{\rm out}\sim20$ analyzing CO outflows with a median velocity of $\sim$15\,km\,s$^{-1}$. However, the outflows in G31 have been observed in SiO, which is a shock tracer likely tracing a  systematically higher outflow component  than CO, which is tracing the bulk outflow gas. Therefore, $V_{\rm jet}/V_{\rm out}$ might be lower, $\sim$10, taking into account that the maximum outflow velocities measured scatter around 30\,km\,s$^{-1}$. Here, we have assumed that $V_{\rm jet}/V_{\rm out}$ is in the range 10--20. Regarding the ratio of ejection to accretion rates,  $\dot M_{\rm jet}/\dot M_{\rm acc}$, theoretical models predict a wide range of values. The X-wind and disk wind models predict that this ratio is in the range 0.1--0.3 (e.g., Tomisaka~\cite{tomisaka98}; Shu et al.~\cite{shu99}; Offner \& Arce~\cite{offner14}). However, ratios as low as 0.01 (or even lower) and as high as 0.5 have been theorized for certain magneto-centrifugally driven wind models (e.g., Nolan et al.~\cite{nolan17}; Sheikhnezami et al.~\cite{sheikhnezami12}). The $\dot M_{\rm jet}/\dot M_{\rm acc}$ ratio is poorly constrained not only theoretically but also observationally (e.g., Calvet et al.~\cite{calvet04}; Cabrit \cite{cabrit07}; Agra-Amboage et al.~\cite{agra09}; Frank et al.~\cite{frank14}, and references therein). Here, we have assumed a ratio in the range 0.1--0.3, which includes the values typically used in the literature, for our calculations. With these ranges of  $V_{\rm jet}/V_{\rm out}$  and $\dot M_{\rm jet}/\dot M_{\rm acc}$, we obtain that $\dot M_{\rm acc} =  0.17$--$1\,\dot M_{\rm out}$. For sources A, B, C, and D, the infall rates are on the order of $\sim$10$^{-2}\,M_\odot$\,yr$^{-1}$ (see Table~\ref{table-infall}), while the accretion rates would be at most (assuming a gas temperature of 50\,K and an abundance SiO range of 10$^{-8}$--10$^{-7}$) on the order of $10^{-4}\,M_\odot$\,yr$^{-1}$ (see Table~\ref{table-infall}).

This difference of about 2  orders of magnitude between $\dot M_{\rm inf}$ and $\dot M_{\rm acc}$ has also been observed in other low- to high-mass protostars (L\'opez-Sepulcre et al.~\cite{lopez10}; see Fig.~16 of Beltr\'an \& de Wit~\cite{beltran16}). Note that the infall rates should be considered as upper limits because the radius at which $V_{\rm inf}$ corresponds might have been overestimated (see Sect.~\ref{sect-indi-out}). On the other hand, for G31, the low values of the accretion rates could be explained in part in terms of SiO missing flux that would underestimate $\dot M_{\rm out}$ and, therefore, $\dot M_{\rm acc}$. To have an estimate of the SiO missing flux, we have compared the total mass of the outflows in the region with the mass estimated by Cesaroni et al.~(\cite{cesa11}) from $^{13}$CO\,(2--1) IRAM 30m and SMA combined data. The total SiO mass is $\sim$1.8\,$M_\odot$ assuming the maximum outflow mass for each individual outflow in Table~\ref{table-outflow}, as used to estimate $\dot M_{\rm acc}$,  while the $^{13}$CO mass is 20\,$M_\odot$ (Cesaroni et al.~\cite{cesa11}). Therefore, the missing flux could account for at least a factor 10. Therefore, taking into account all the uncertainties in the estimates of $\dot M_{\rm inf}$ and $\dot M_{\rm acc}$, the ratio $\dot M_{\rm inf}/\dot M_{\rm acc}$ could range from a few to 100. We believe that most likely the true value lies in between these two extremes and thus conclude that we cannot discard the possibility that the efficiency of the central star and the jet in removing material from the accretion disk is lower than expected, as suggested by Beltr\'an \& de Wit~(\cite{beltran16}). The infalling material could pile up in the disk, which would increase its mass and ultimately  become  gravitationally unstable and fragment (e.g., Peters et al.~\cite{peters11}; Oliva \& Kuiper~\cite{oliva20}). This could lead to episodic accretion bursts onto the central object similar to those observed toward high-mass sources like  S255 NIR 3 (Caratti o Garatti et al.~\cite{caratti16}), NGC\,6334 I-MM1 (Hunter et al.~\cite{hunter17}, \cite{hunter18}, \cite{hunter21}), G358.93$-$00.03 (Sugiyama et al.~\cite{sugiyama19}), and W51-North (Goddi et al.~\cite{goddi20}), and theoretically quantified by e.g., Meyer et al.~(\cite{meyer17}, \cite{meyer19}) and Oliva \& Kuiper~(\cite{oliva20}).

\subsection{Outflow feedback}

Taking into account that all the sources embedded in the {\it Main} core of G31 are driving outflows, we investigated the impact of such outflows on the G31 core, and in particular, we studied whether the outflow feedback is sufficient to sustain the turbulence in the $Main$ core and also to disrupt the core itself. Because the extended emission of the outflows as traced in SiO with ALMA is significantly affected by missing flux as seen in the previous section, for our calculations we decided to use the properties of the outflows estimated by Cesaroni et al.~(\cite{cesa11}) from $^{13}$CO\,(2--1) IRAM 30m and SMA combined data. The resolution of these observations ($\sim$1$\farcs7$) is not high enough to resolve the individual outflows, but this is not important in this case because we are interested in the total properties (momentum, momentum rate, energy, energy rate) of the outflows.

Following Stanke \& Williams~(\cite{stanke07}), we first estimated whether the energy and momentum injected by the outflows were enough to sustain the turbulence in the core, which should decay on a sound crossing time, $t_{\rm c_s}$. Assuming a temperature of 250\,K, which is the mean value of the temperature in the core estimated by Beltr\'an et al.~(\cite{beltran18}), and the size of $\sim$8000\,au, the sound speed, $c_{\rm s}$, is $\sim$1\,km\,s$^{-1}$, and $t_{\rm c_s}$ is $\sim$4$\times10^4$\,yr. Following Stanke \& Williams~(\cite{stanke07}), the turbulent energy can be estimated as $E_{\rm turb}=3/2\,M_{\rm core}\,\sigma_{\rm 1D}^2$, where $M_{\rm core}$ is the mass of the {\it Main} core, which is $\sim$70\,$M_\odot$ (Cesaroni et al.~\cite{cesa19}), and $\sigma_{\rm 1D}$ is the one-dimensional velocity dispersion, estimated to be 2.1\,km\,s$^{-1}$\ by Beltr\'an et al.~(\cite{beltran19}). This estimate is essentially  turbulent velocity dispersion, because thermal and rotation contributions to the line broadening are negligible as discussed by Beltr\'an et al.~(\cite{beltran19}). Therefore, $E_{\rm turb}\sim9\times10^{45}$\,erg. The rate at which the energy is lost due to decay of turbulence is then calculated as $L_{\rm turb} = E_{\rm turb}/t_{\rm c_s}$, and it is $\sim$2\,$L_\odot$ for G31.  The total mechanical luminosity of the outflows, $L_{\rm mech}$, is $\sim$55\,$L_\odot$, which is $\sim$28 times larger than $L_{\rm turb}$. These numbers are consistent if $\sim$4\% of the outflow momentum is transferred to the core and the remainder is dissipated by shocks or just leaves the core. 

The rate $L_{\rm gain}$ at which the gas cloud gains energy, taking into account radiative losses, can be estimated following Stanke \& Williams~(\cite{stanke07}) as $L_{\rm gain}=\sqrt{ 3}/2\,\dot P_{\rm out}\sigma_{\rm 1D}$, where $\dot P_{\rm out}$ is the total outflow momentum rate, estimated to be 0.057\,$M_\odot$\,km\,s$^{-1}$\,yr$^{-1}$ by Cesaroni et al.~(\cite{cesa11}). The energy gain rate is $L_{\rm gain}$$\sim8\,L_\odot$, similar to $L_{\rm turb}$, which suggests that the molecular outflows in G31 can inject enough  energy and momentum to sustain the turbulence in the core.

Following Rivilla et al.~(\cite{rivilla13}), we evaluated whether the feedback of the outflows in the dense and compact {\it Main} core was enough to disrupt it. To do this, we compared the energy of the outflows, $E_{\rm out}$ with the gravitational potential energy of the core, $\lvert E_{\rm grav}\rvert$, which can be calculated as $G\,M_{\rm core}^2/R_{\rm core}$ where $R_{\rm core}$ is the radius of the core estimated to be $\sim$1$\farcs$1 ($\sim$4000\,au) by Beltr\'an et al.~(\cite{beltran18}). Since the outflows are still deeply embedded in the core, especially outflows A and B, their energy is likely to be deposited locally and, therefore, we do not expect considerable energy leakage. Cesaroni et al.~(\cite{cesa11}) have calculated $E_{\rm out}$ $\sim$3$\times10^{46}$\,erg, while $\lvert E_{\rm grav}\rvert$ is $\sim$2$\times10^{46}$\,erg.  Since the {\it Main} core in G31 is not in equilibrium but is apparently collapsing, to prevent further global collapse of the core, the outflows should counteract not only the gravitational energy of the core but also the kinetic energy of the infalling material, $E_{\rm kin}$. Virial theorem considerations suggest $E_{\rm kin} \sim 1/2\,\lvert E_{\rm grav}\rvert \sim 1\times10^{46}$\,erg. Since the energy of the outflows is comparable to the sum of $\lvert E_{\rm grav}\rvert$ and $E_{\rm kin}$, the outflows in G31 could, in principle, prevent further global collapse of the core. We thus do not expect further star formation to occur in G31.

\section{Conclusions}

We carried out ALMA line observations at 1.4\,mm of the high-mass star-forming region G31 at an angular resolution of $\sim$0$\farcs$09 ($\sim$340\,au). The goal of these observations was to study the kinematics of the four sources, A, B, C, and D, embedded in the {\it Main} core of G31. In particular, our aim was to better investigate at higher angular resolution the outflows and the collapse in the core previously discovered by Beltr\'an et al.~(\cite{beltran18}). 

The observations have revealed that the four sources are undergoing collapse, as suggested by the presence of red-shifted absorption in CH$_3$CN and H$_2$CO. The infall velocities estimated from H$_2$CO are $\sim$2--3\,km\,s$^{-1}$\, and the infall rates, which should be taken as upper limits, are on the order of  $\sim$10$^{-2}$\,$M_\odot$\,yr$^{-1}$. The similarity between the free-fall times of the individual sources A to D and the accretion timescale of the {\it Main} core, estimated from the individual infall rates, suggests that the embedded sources evolve dynamically together with the {\it Main} core, and that the forming stars can tap into the entire reservoir of material in the core. The fact that the accretion timescale of the {\it Main} core is an order of magnitude smaller than its rotation timescale confirms that rotating toroids, such as the G31 {\it Main} core,  are non-equilibrium, transient collapsing structures, constantly replenished with fresh material from a large-scale reservoir. 

The infall velocity estimated from CH$_3$CN and isotopologues, and H$_2$CO for source B, C, and D, increases with the upper level energy of the transition, which suggests that infall could be accelerating within the sources. For source A, the presence of a second  absorption component complicates the estimation of the infall velocity. 

The SiO observations indicate the presence of at least six outflows in the G31 star-forming region, and suggest that each of the four sources embedded in the {\it Main} core drives a molecular outflow. This confirms that the sources are still actively accreting material. The outflow rates are on the order of $\sim$10$^{-5}$--10$^{-4}$\,$M_\odot$\,yr$^{-1}$, depending on the SiO abundance. The outflows with the highest mass loss rates are outflow E--W, associated with source D, and outflow SW, driven by source A, while the most embedded one is outflow SE, powered by source B. The latter outflow is also the one with the smallest velocities and size of the lobes. The four outflows are likely associated with H$_2$O maser emission. 

The mass accretion rates onto the individual sources, estimated from the highest value of the mass loss rate, are on the order of 10$^{-4}$\,$M_\odot$\,yr$^{-1}$, about 2 orders of magnitude smaller than the infall rates. This difference between $\dot M_{\rm acc}$ and $\dot M_{\rm inf}$ may be partly due to filtering of the extended outflow emission, uncertainties in the ejection to accretion ratio, or an overestimation of the infall rates, but it could also be real and suggest inefficient removal of the disk material from the central star and the jet, which could lead to gravitational instabilities and episodic accretion events.
 
Our study indicates that infall and outflows are simultaneously present in all four sources embedded in the high-mass star-forming core G31.41+0.31. This indicates that these sources are still actively accreting and have not reached their final mass yet, and confirms that also in the high-mass regime, accretion seems to be commonly associated with outflow activity.  The energy and momentum available in the outflows have the potential to inject significant turbulence in the {\it Main} core and to eventually disrupt it or prevent its further collapse. This suggests that the number of sources associated with the small protocluster might remain unchanged.

\begin{acknowledgements}

We thank the anonymous referee for useful comments that have improved the manuscript. This paper makes use of the following ALMA data: ADS/JAO.ALMA\#2013.1.00489.S and ADS/JAO.ALMA\#2016.1.00223.S. ALMA is a partnership of ESO (representing its member states), NSF (USA) and NINS (Japan), together with NRC (Canada) and NSC and ASIAA (Taiwan), in cooperation with the Republic of Chile. The Joint ALMA Observatory is operated by ESO, AUI/NRAO and NAOJ.   The National Radio
Astronomy Observatory is a facility of the National Science Foundation operated under cooperative agreement by Associated Universities, Inc. V.M.R.\ acknowledges support from the Comunidad de Madrid through the Atracci\'on de Talento Investigador Modalidad 1 (Doctores con experiencia) Grant (COOL: Cosmic Origins Of Life; 2019-T1/TIC-15379). H.B.\ acknowledges support from the European Research Council under the Horizon 2020 Framework Program via the ERC Consolidator Grant CSF-648405. H.B.\ also acknowledges support from the Deutsche Forschungsgemeinschaft in the Collaborative Research Center (SFB881) ``The Milky Way System'' (subproject B1). R.K.\ acknowledges financial support via the Emmy Noether and Heisenberg Research Grants funded by the German Research Foundation (DFG) under grant no.~KU 2849/3 and 2849/9. M.S.N.K.\ acknowledges the support from FCT - Fundac\~{a}o para a Ci\^{e}ncia e a Tecnologia through Investigador contracts and exploratory project (IF/00956/2015/CP1273/CT0002) 
\end{acknowledgements}

\appendix

\section{Mass infall rate}

The infall rate in a spherically symmetric clump of radius $R_{\rm c}$ is given by the expression
\begin{equation}
    \dot M_{\rm inf} = \Omega\, V_{\rm inf}(R_{\rm c})\, \mu\, m_{\rm H} n_{\rm H_2}(R_{\rm c}) R_{\rm c}^2 \label{einf}
\end{equation}
where $\Omega$ is the solid angle subtended by the clump center over which infall occurs, $V_{\rm inf}$ the infall velocity, $m_{\rm H}$ the mass of the H atom, $\mu=2.8$ the mean molecular weight, and 
 $n_{\rm H_2}$ the number density of the H$_2$ molecules. However, it is more convenient to replace the density at the clump radius with the mean density inside that radius, $\langle n_{\rm H_2}\rangle$, as the latter is the quantity that is obtained from the observations. For this purpose, we assume that the clump is undergoing free-fall collapse, $V_{\rm inf} \propto r^{-1/2}$, with $r$ radius from the clump center. Mass conservation, i.e. $\dot M_{\rm inf}={\rm const.}$, implies that for any $r$ the following condition must be satisfied
\begin{equation}
    V_{\rm inf}(r)\, n_{\rm H_2}(r) \, r^2 = V_{\rm inf}(R_{\rm c})\, n_{\rm H_2}(R_{\rm c}) \,R_{\rm c}^2
\end{equation}
or, equivalently,
\begin{equation}
    r^{-\frac{1}{2}}\, n_{\rm H_2}(r)\, r^2 = R_{\rm c}^{-\frac{1}{2}}\, n_{\rm H_2}(R_{\rm c})\, R_{\rm c}^2
\end{equation}
which implies
\begin{equation}
    n_{\rm H_2}(r) = n_{\rm H_2}(R_{\rm c}) \left(\frac{r}{R_{\rm c}}\right)^{-\frac{3}{2}}.
\end{equation}
One can now compute the mean density inside the clump:
\begin{eqnarray}
    \langle n_{\rm H_2} \rangle & = & 
    \frac{3}{4\pi R_{\rm c}^3} \int_0^{R_{\rm c}} n_{\rm H_2}(r)\, 4\pi r^2 \,{\rm d}r \nonumber \\
    & = & 3 \, n_{\rm H_2}(R_{\rm c}) \int_0^1 \left(\frac{r}{R_{\rm c}}\right)^{\frac{1}{2}} \, {\rm d}\left(\frac{r}{R_{\rm c}}\right) \\
    & = & 2\,n_{\rm H_2}(R_{\rm c}). \nonumber
\end{eqnarray}
Finally, Eq.~(\ref{einf}) can be written in the form
\begin{equation}
    \dot M_{\rm inf} = \frac{\Omega}{2}\, V_{\rm inf}(R_{\rm c})\, \mu\, m_{\rm H} \langle n_{\rm H_2}\rangle R_{\rm c}^2
\end{equation}

\begin{thebibliography}{}

\bibitem[2009]{agra09}
Agra-Amboage, V., Dougados, C., Cabrit, S., Garcia, P.\ J.\ V., \& Ferruit, P.\
2009, A\&A, 493, 1029


\bibitem[2008]{arce08}
Arce, H.\ G.,  Santiago-Garc\'{i}a, J., J{\o}rgensen, J.\ K., Tafalla, M., \& Bachiller, R.\ 2008, ApJL, 681, 21
 
\bibitem[2007]{arce07}
Arce, H.\ G., Shepherd, D., Gueth, F.\ et al.\ 2007, in Protostars and Planets V, eds.\ B.\ Reipurth, D.\ Jewitt, \& K.\ Keil, University of Arizona Press, Tucson, 245 

 

\bibitem[2011]{beltran11} 
Beltr\'an, M.\ T.\ 2011, in Computational Star Formation, Proceedings of the International Astronomical Union, IAU Symposium, Vol.\ 270, 33
 
\bibitem[2020]{beltran20} 
Beltr\'an, M.\ T.\ 2020, in Perspective of the Star Formation Newsletter n.\ 329, arXiv:2005.06912 (https://arxiv.org/abs/2005.06912)

\bibitem[2016]{beltran16} 
Beltr\'an, M.\ T., \& de Wit, W.\ J.\ 2016, A\&ARv 24, 6

\bibitem[2011a]{beltran11a} 
Beltr\'an, M.\ T., Cesaroni, R., Neri, R., Codella, C.\ 2011a, A\&A, 525, A151

\bibitem[2004]{beltran04} 
Beltr\'an, M.\ T., Cesaroni, R., Neri, R., Codella, C., Furuya, R.\ S., Testi, L., \& Olmi, O.\ 2004, ApJ, 601, L190 

\bibitem[2005]{beltran05} 
Beltr\'an, M.\ T., Cesaroni, R., Neri, R., Codella, C., Furuya, R.\ S., Testi, L., \& Olmi, O.\ 2005, A\&A, 435, 901 

\bibitem[2018]{beltran18} 
Beltr\'an, M.\ T., Cesaroni, R., Rivilla, V.\ M., S\'anchez-Monge, \'A.\ et al.\  2018, A\&A, 615, A141 

\bibitem[2011b]{beltran11b} 
Beltr\'an, M.\ T., Cesaroni, R., Zhang, Q., Galv\'an-Madrid, R.\ et al.\ 2011b, A\&A, 532, A91 


\bibitem[2019]{beltran19} 
Beltr\'an, M.\ T., Padovani, M., Girart, J.\ M., Galli, D., et al.\  2019, A\&A, 630, A54 

\bibitem[2021]{beltran21} 
Beltr\'an, M.\ T., Rivilla, V.\ M., Cesaroni, R., Maud, L.\ T.\ et al.\  2021, A\&A, 648, A100  

\bibitem[2002]{beuther02}
Beuther, H., Schilke, P., Sridharan, T.\ et al.\ 2002, A\&A, 383, 892


\bibitem[1995]{briggs95}
Briggs, D.\ 1995,  PhD Thesis, New Mexico Inst.\ Mining \& Tech.\

\bibitem[2007]{cabrit07}
Cabrit, S.\ 2007, in IAU Symp.\ 243, Star-Disk Interaction in Young Stars, ed.\
J.\ Bouvie \& I.\ Appenzeller (Cambridge: Cambridge Univ.\ Press), 203


\bibitem[2004]{calvet04}
Calvet, N., Muzerolle, J., Brice\~no, C.\ et al.\ 2004, AJ, 128, 1294

\bibitem[2016]{caratti16}
Caratti o Garatti, A.,  Stecklum, B., Weigelt, G.\ et al.\ 2016, A\&A, 589, L4

\bibitem[2019]{cesa19}
Cesaroni, R.\ 2019, A\&A 631, A65 

\bibitem[2011]{cesa11}
Cesaroni, R., Beltr\'an, M.\ T., Zhang, Q., Beuther, H., \& Fallscheer, C.\ 2011, A\&A, 533, A73


\bibitem[2010]{cesa10}
Cesaroni, R., Hofner, P., Araya, E., \& Kurtz, S.\ 2010, A\&A, 590, A50 (C10)

\bibitem[2017]{cesa17}
Cesaroni, R., S\'anchez-Monge, \'A., Beltr\'an, M.\ T., Johnston, K.\ G., et al.\ 2017, A\&A, 602, A59

\bibitem[2005]{chandler05}
Chandler, C.\ J., Brogan, C.\ L., \& Shirley, Y.\ L.\ 2005, ApJ, 632, 371




\bibitem[2009]{codella09}
Codella, C., Benedettini, M., Beltr\'an, M.\ T.\ et al.\ 2009, A\&A, 507, L25 

\bibitem[2013]{codella13}
Codella, C., Beltr\'an, M.\ T., Cesaroni, R., Moscadelli, L.\ et al.\ 2013,
A\&A, 550, A81





\bibitem[2014]{frank14}
Frank, A., Ray, T.\ P., Cabrit, S.\ et al.\ 2014, in Protostars and Planets VI, eds.\ H.\ Beuther, R.\ S. Klessen, C.\ P.\ Dullemond, \& Th.\ Henning, University of Arizona Press, 451 


\bibitem[2021]{gieser21}
Gieser, C., Beuther, H., Semenov, D.\ 2021, A\&A, 648, A66

\bibitem[2019]{gieser19}
Gieser, C., Semenov, D., Beuther, H.\ et al.\ 2019, A\&A, 631, A142

\bibitem[2009]{girart09}
Girart, J.\ M., Beltr\'an, M.\ T., Zhang, Q., Rao, R., \& Estalella, R.\ 2009, Science, 324, 1408


\bibitem[2020]{goddi20}
Goddi, C., Ginsburg, A., Maud, L.\ T., Zhang, Q., \& Zapata, L.\ A.\ 2010, ApJ, 905, 25



\bibitem[1986]{ho86}
Ho, P.\ T.\ P., \& Haschick, A.\ D.\ 1986, ApJ, 304, 501


\bibitem[2021]{hunter21}
Hunter, T.\ R., Brogan, C.\ L., De Buizer, J.\ M.\ et al.\ 2021, ApJ, 912, L17

\bibitem[2017]{hunter17}
Hunter, T.\ R., Brogan, C.\ L., MacLeod, G.\ et al.\ 2017, ApJ, 837, L29

\bibitem[2018]{hunter18}
Hunter, T.\ R., Brogan, C.\ L., MacLeod, G.\ C.\ et al.\ 2018, ApJ, 854, 170

\bibitem[2019]{immer19}
Immer, K., Li, J., Quiroga-Nu\~nez, L.\ H., et al.\ 2019, A\&A, 632, A123


\bibitem[1988]{keto88}
Keto, E.\ R., Ho, P.\ T.\ P., \& Haschick, A.\ D.\ 1988, ApJ, 324, 920

\bibitem[2015]{krumholz15}
Krumholz, M.\ R.\ 2015, in Very Massive Stars in the Local Universe, Astrophysics and Space Science Library, 412, 43




\bibitem[2018]{kuiper18}
Kuiper, R., \& Hosokawa, T.\ 2018, A\&A, 616, A101



\bibitem[2010]{lopez10}
L\'opez-Sepulcre, A., Cesaroni, R., \& Walmsley, C.\ 2010, A\&A, 517, A66 




\bibitem[1977]{mathis77}
Mathis, J.\ S., Rumpl, W., \& Nordsieck, K.\ H.\ 1977, ApJ,217, 425




\bibitem[2019]{meyer19}
Meyer, D.\ M.-A., Vorobyov, E.\ I., Elbakyan, V.\ G.\ et al.\ 2019, MNRAS, 482, 5459 

\bibitem[2017]{meyer17}
Meyer, D.\ M.-A., Vorobyov, E.\ I., Kuiper, R., \& Kley, W.\ 2017, MNRAS, 464, L90 


\bibitem[2021]{mosca21}
Moscadelli, L.,  Cesaroni, R.\ Beltr\'an, M.\ T., \& Rivilla, V.\ M.\ 2021, A\&A, 650, A142

\bibitem[2013]{mosca13}
Moscadelli, L., Li, J.\ J., Cesaroni, R., Sanna, A., Xu, Y., \& Zhang, Q.\ 2013, A\&A, 549, A122

\bibitem[2018]{mosca18}
Moscadelli, L., Rivilla, V.\ M., Cesaroni, R.\ Beltr\'an, M.\ T.\  et al.\ 2018, A\&A, 616, A66


\bibitem[2018]{motte18}
Motte, F., Bontemps, S., \& Louvet, F.\ 2018, ARA\&A, 56, 41


\bibitem[2017]{nolan17}
Nolan, C.\ A., Salmeron, R., Federrath, C.\ et al.\ 2017, MNRAS, 471, 1488


\bibitem[2014]{offner14}
Offner,  S.\ S.\ R., \& Arce, H.\ G.\ 2014, ApJ, 784, 61

\bibitem[2020]{oliva20}
Oliva, G.\ A., \& Kuiper, R.\ 2020, A\&A, 644, A41

\bibitem[1996]{olmi96}
Olmi, L., Cesaroni, R., \& Walmsley, C.\ M.\ 1996, A\&A, 307, 599

\bibitem[2009]{osorio09}
Osorio, M., Anglada, G., Lizano, S., \& D’Alessio, P.\ 2009, ApJ, 694, 29

\bibitem[1994]{ossenkopf94}
Ossenkopf, V., \& Henning, Th.\ 1994, A\&A, 291, 943

\bibitem[2017]{palau17}
Palau, A., Walsh, C., S\'anchez-Monge, \'A.\ et al.\ 2017, MNRAS, 467, 2723 



\bibitem[2011]{peters11}
Peters, T., Banerjee, R., Klessen, R.\ S., \& Mac Low, M.-M.\ 2011, ApJ, 729, 72





\bibitem[2017]{rivilla17}
Rivilla, V.\ M., Beltr\'an, M.\ T., Cesaroni, R., et al.\ 2017, A\&A, 598, A59

\bibitem[2013]{rivilla13}
Rivilla, V.\ M., Mart{\'i}n-Pintado, J., Sanz-Forcada, J.\ et al.\ 2013, MNRAS, 434, 2313 



\bibitem[2018]{sanchez-monge18}
S\'anchez-Monge, \'A., Schilke, P., Ginsburg, A., Cesaroni, R., \& Schmiedeke, A.\ 2018, A\&A, 609, A101


\bibitem[2012]{sheikhnezami12}
Sheikhnezami, S., Fendt, Ch., Porth, O.\ et al.\ 2012, ApJ, 757, 65

\bibitem[1996]{shepherd96}
Shepherd, D.\ S., \& Churchwell, E.\ 1996, ApJ, 472, 225

\bibitem[1987]{shu87}
Shu, F.\ H., Adams, F.\ C., \& Lizano, S.\ 1987, ARA\&A, 25, 23


\bibitem[1999]{shu99}
Shu, F., Allen, A., Shang, H., Ostriker, E., \& Li, Z.-Y.\ 1999, in NATO Advanced
Science Institutes (ASI) Series C, eds.\ C.\ Lada, \& N.\ Kylafis, vol. 540, 193

\bibitem[2007]{stanke07}
Stanke, T., \& Williams, J.\ P.\ 2007, AJ, 133, 1307

\bibitem[2019]{sugiyama19}
Sugiyama, K.,  Saito, Y., Yonekura, Y., \& Momose, M.\ 2019, The Astronomer’s Telegram, 12446, 1


\bibitem[1998]{tomisaka98}
Tomisaka, K.\ 1998, ApJ, 502, L163

\bibitem[2019]{vandertak19}
van der Tak, F.\ F.\ S., Shipman, R.\ F., Jacq, T.\ et al.\ 2019, A\&A, 625, A103

\bibitem[2002]{wyrowski02}
Wyrowski, F., Gibb, A.\ G., Mundy, L.\ G.\ 2002, in Hot Star Workshop III: The Earliest Stages of Massive Star Birth. ASP Conference Proceedings, Vol. 267, 43

\bibitem[2012]{wyrowski12}
Wyrowski, F., G\"usten, R., Menten, K.\ M.\ et al.\  2012, A\&A, 542, L15

\bibitem[1999]{wyrowski99}
Wyrowski, F., Schilke, P., \& Walmsley, C.\ M.\ 1999, A\&A, 341, 882


\end{thebibliography}
\end{document}